\documentclass[preprint,preprintnumbers,amsmath,amssymb]{revtex4}
\usepackage{graphicx}
\usepackage{graphics}
\usepackage{chngcntr}
\usepackage{bm}
\usepackage{epsfig}
\begin{document}
	
	\title{Lepton mass effects in exclusive semileptonic $B_c$-meson decays}

	\author{Lopamudra Nayak$^{1}$\footnote{email address:lopalmn95@gmail.com}, Sonali Patnaik$^{1}$, P. C. Dash$^{1}$, Susmita Kar$^{2}$, N. Barik$^{3}$}
	\affiliation{$^{1}$ Department of Physics, Siksha $'O'$ Anusandhan University, Bhubaneswar-751030,\\$^{2}$ Department of Physics, North Orissa University, Baripada-757003,\\ 
		$^{3}$ Department of Physics, Utkal University, Bhubaneswar-751004.}

	\begin{abstract}
		In this work, we discuss exclusive semileptonic $B_c$-meson decays: $B_c\to \eta_c(J/\psi)l\nu$ and $B_c\to D(D^*)l\nu$ in the framework of the relativistic independent quark(RIQ) model based on an average flavor independent confining potential in equally mixed scalar-vector harmonic form. We calculate the invariant form factors representing decay amplitudes from the overlapping integrals of meson wave functions derivable in the RIQ model. To evaluate the lepton mass effects in the semileptonic decays, we first study the $q^2$-dependence of the form factors in the accessible kinematic range of $q^2$ involved in the decay process in its $e^-$ and $\tau^-$ mode separately. Similar studies on helicity amplitudes, $q^2-$spectra for different helicity contributions, and total $q^2$-spectra for each decay process are carried out separately in their $e^-$ and $\tau^-$ modes. We predict the decay rates/ branching fractions, forward-backward asymmetry, and the asymmetry parameter in reasonable agreement with other model predictions, which can hopefully be tested in future experiments at the Tevatron and LHC. We also predict the observable $'R'$ which corresponds to the ratio of branching fractions for the decay process in its $e^-$ mode to its corresponding value in the $\tau^-$ mode. Our results are comparable to another standard model(SM) predictions which highlight the failure of the lepton flavor universality hinting at new physics beyond SM for the explanation of the observed deviation of observable $'R'$ value from the corresponding SM predictions.
			\end{abstract}
	\maketitle
	\section{Introduction}
	\noindent  $B_c$-meson is the lowest bound state of two heavy quarks (charmed and bottom) with open(explicit) flavors. As far as bound state characteristics are concerned, $B_c$-meson is quite similar to charmonium ($c\bar{c}$ bound state) and bottomonium ($b\bar{b}$ bound state). The double heavy quarkonia with hidden(implicit) flavors, can decay strongly and electromagnetically whereas $B_c$-meson decays weakly since it lies below the $B\bar{D}$ threshold. That makes it an ideal system for the study of heavy quark dynamics. As such $B_c$-meson has a long lifetime. Both its constituent quarks$(b,c)$ being heavy they can decay yielding a large number of $B_c$-meson weak channels with sizeable branching fractions. Although it lies intermediate in mass and size between charmonium and bottomonium family where heavy quark interactions are understood rather well, many aspects of the weak interaction in the $B_c$-sector remain obscure due to lack of adequate data.\\
	Ever since its discovery at Fermilab by the CDF Collaboration\cite{A1}, a series of experimental probes have been made yielding only a few observed data  so far. The $B_c$-meson lifetime has been measured \cite{A2,A3,A4,A5} using the decay channels: $B_c^{\pm}\to J/\psi l^{\pm}\nu_l$ and $B_c^\pm\to J/\psi \pi^\pm$. A more precise measurement of its lifetime and mass by LHCb\cite{A6} yields $\tau_{B_c}={0.51^{+0.18}_{-0.16}}(stat.)\pm 0.03(syst.)ps$ and $M_{B_c}=6.40\pm0.39\pm 0.13\ GeV$, respectively, using the decay mode $B_c \to J/\psi l\nu_l X $, where $X$ denotes any possible additional particle in the final state. The mass of the excited state: $B_c(2S)$ has also been observed by the ATLAS Collaboration \cite{A7} in their analysis of the decay channel $B_c(2S)\to B_c(1S)\pi^-\pi^+$ by using $4.9 fb^{-1}$ of $7\ TeV$ and $19.2fb^{-1}$ of 8\ TeV pp collision data. The current Run II at Tevatron \cite{A8,A9} and upcoming Run III at the CERN LHC as well as the $e^+ e^-$ experiment by Belle-II are designed to boost the heavy flavor physics measurement scenario in near future. The dedicated detectors at $B$ TeV and LHCb specially designed to enhance the event accumulation rates are expected to provide high statistics $B_c$-events at the rate of $~10^{10}$ events per year.\\
	Among various weak decays of $B_c$-meson, it is semileptonic(s.l.) decay mode is significant, since $B_c$-meson is first observed by CDF Collaboration \cite{A1} in their analysis of the decay mode $B_c\to J/\psi l\nu_l$ with $J/\psi $ decaying into muon pair. Besides extracting the accurate value of Cabbibo-Kobayashi-Mashakawa(CKM) matrix elements, the study of s.l. decays help in examining the universality of coupling of three charged leptons in the electroweak interactions. Thus, it provides a powerful tool for testing the SM and searching effects of physics beyond SM. Due to their simple theoretical description via tree-level processes in the SM, the analysis of semileptonic decays helps in separating the effects of strong interaction from that of the weak interaction into a set of Lorentz invariant form factors. The study of s.l. decay processes, therefore, reduces to a calculation of relevant form factors in a suitable phenomenological model framework.\\\\
	The transition form factors parametrizing s.l. decay amplitudes are evaluated from the overlapping integrals of meson wave functions obtainable in different theoretical approaches. Some of them include the potential model approach\cite{A10}, the Bethe Salpeture approach \cite{A11,A12}, relativistic constituent quark model on the light front\cite{A13,A14}, three-point sum rule of QCD, and non-relativistic QCD(NRQCD)\cite{A15,A16,A17}, relativistic quark model based on the quasi-potential approach\cite{A18}, non-relativistic quark model approach\cite{A19}, the Baur, Stech-Wirbel framework\cite{A20}, perturbative QCD(PQCD) approach\cite{A21,A22}, the covariant confined quark model approach\cite{A23,A24,A25,A26} and the lattice QCD approach\cite{A27}. The relativistic independent quark(RIQ) model, developed by our group and applied extensively in the description of wide-ranging hadronic phenomena including the static properties of hadrons \cite{A28}, and their decay properties such as the radiative, weak radiative, rare radiative\cite{A29}; leptonic, weak leptonic, radiative leptonic\cite{A30} and non-leptonic\cite{A31} decays, has also been applied in analyzing the s.l. decays of heavily flavored mesons\cite{A32,A33,A34,A35}.\\
	It may be noted here that our analysis based on the vanishing lepton mass limit may be considered reasonable in the description of semileptonic decay modes\cite{A32,A33,A34,A35} where only the 3-vector (or space component) hadronic current form factors contribute to the decay amplitudes. The scalar-(time component) hadronic current form factor is not accessible in such a description. However, such an approach is not applicable in describing the semi-tauonic decay modes where both space and time component hadronic current form factors contribute to the decay amplitudes. A series of measurements for semi-tauonic decays: $B\to D(D^*)\tau \nu_\tau$ have been reported by BaBar\cite{A36,A37}, Belle\cite{A38,A39,A40}, and LHCb\cite{A41,A42}. Recently, the measurement for $B_c\to J/\psi \tau \nu_{\tau}$ has also been reported by LHCb Collaboration\cite{A43}.\\
	One of the most puzzling issues in recent years in flavor physics has been the observed deviation of the observable: $	R(J/{\psi})=\frac{{\cal B}({B_c^+}\to J/{\psi}{\tau}^+{\nu}_{\tau})}{{\cal B}({B_c^+}\to J/{\psi} {\mu}^+{\nu}_{\mu})}$ which corresponds to the ratio of branching fractions, over the SM predictions. The measurement by LHCb Collaboration\cite{A43} yields $R(J/{\psi})=0.71\pm 0.17\pm0.18$ whereas the central values of the current SM predictions are in the range of $0.25-0.28$ which is about $2\sigma$ lower. Here the spread of SM prediction is due to the choice of modeling approach \cite{A44,A45,A46,A47} for the form factors. For definiteness, the most recent result\cite{A48} used as SM values is $R(J/\psi)=0.283\pm0.048$. This anomaly between the observed data and corresponding SM predictions hints at the failure of lepton flavor universality. Attempts have been made to explain the discrepancy via the possible extension of the SM that involves an enhanced weak coupling to three charged leptons and quarks in electroweak interactions.\\
	The s.l. decay modes: $B_c\to \eta_c(J/\psi)l\nu_l$ are induced by the quark level transition: $b\to c l \nu_l$. Identical to these modes, $B_c\to D(D^*)l\nu_l$ are induced by $b\to ul\nu_l$ at the quark level. The kinematic range of $q^2$ for the former class of modes is $0\leq {q^2}\leq10\ GeV^2$, whereas the $q^2$ range for the latter type decay modes is $0\leq {q^2}\leq18\ GeV^2$. In the parent($B_c$-) meson rest frame, the maximum recoil momenta of the final state charmonium($\eta_c, J/\psi$) or charm mesons($D, D^*$) can therefore be estimated to be in the same order of magnitude as their masses. With these kinematic constraints, it is interesting to analyze the decay modes:  $B_c\to \eta_c(J/\psi)l\nu_l$ and  $B_c\to D(D^*)l\nu_l$. A recent analysis of these decay modes by different theoretical approaches has reported their standard model predictions\cite{A25,A46,A49} on the observables: $R(D), R(D^*), R(\eta_c)$, and $R(J/\psi)$.\\
	It is worthwhile to note here following few points:(1) In our RIQ model, the relevant form factors are evaluated in the full kinematic range of momentum transfer squared $q^2$, which makes our prediction more accurate. In some of the theoretical approaches cited above, the form factors are determined first, with an endpoint normalization at minimum $q^2$(maximum recoil point) or maximum $q^2$(minimum recoil point). Then they are phenomenologically extrapolated to the whole physical region using some  monopoles/ dipoles /Gaussian ansatz which makes form factor estimation less reliable. (2) We would evaluate the relevant hadronic current form factors (vector as well as scalar parts) to analyze all possible s.l. $B_c$-meson decay modes induced by $b\to cl\nu_l$ and $b \to u l \nu_l$ transition at the quark level. In doing so we intend to study the lepton mass effects in the s.l. $B_c$-decays and predict the observables $R$ in comparison with other SM predictions. (3) We shall update some input hadronic parameters in our calculation according to the Particle Data Group\cite{A50}.\\
	The paper is organized as follows. In Section II we discuss the general formalism and kinematics for $B_c$-meson s.l. decays. We provide a brief description of the RIQ model framework and extract model expression of transition form factors in Section III. Section IV contains our numerical results and discussion. In Section V we briefly summarize our results.

	\section{General Formalism and Kinematics}
The invariant matrix element for exclusive s.l. $B_c$-decays: $B_c\to \eta_c(J/\psi)l^-\bar{\nu}_l$ and $B_c\to D(D^*)l^-\bar{\nu}_l$ is written in the general form :
	\begin{equation}
		{\cal M}(p,k,k_l,k_\nu)={\frac{\cal G_F}{\sqrt{2}}}V_{bq^{'}}{\cal H}_\mu(p,k) {\cal L}^\mu(k_l,k_\nu) 
	\end{equation}
	where ${\cal G}_F$ is the effective fermi coupling constant, $V_{bq^{'}}$ 
	is the  relevant CKM parameter, ${\cal L}^\mu$ and ${\cal H}_\mu$ are leptonic and hadronic current, respectively given by
	 
	\begin{eqnarray}
		{\cal L}^\mu(k_l,k_\nu)=&&\bar u(\vec{k_l})\gamma ^\mu (1-\gamma_5)v(\vec{k_\nu})\nonumber\\
		{\cal H}_\mu(p,k) =&&\langle X(\vec{k},S_X)|{J^h_\mu(0)}|B_c(\vec{p},S_{B_c})\rangle
	\end{eqnarray}
	Here $J^h_\mu=V_\mu-A_\mu$ is the vector-axial vector current; $q^{'}=c,u$. We take $(p,k)$ as four momenta of the parent($B_c$) and daughter($X$) meson with their respective spin state: $S_{B_c}$ and $S_X$ mass: M and m. $k_l$ and $k_\nu$ are the four momenta of lepton and lepton neutrino, respectively. $q=p-k=k_l+k_\nu$ represents the four-momentum transfer.\\
	The hadronic amplitudes are covariantly expanded in terms of a set of Lorentz invariant form factors.\\
	For $(0^- \to 0^-)$ type transitions, it is defined as 
	\begin{equation}
		{\cal H}_\mu(B_c\to(\bar{c}c/\bar{u}c)_{S=0}) =(p+k)_\mu F_+(q^2)+q_\mu F_-(q^2)
	\end{equation}
	For $(0^-\to 1^-)$ type transitions, the expansion is 
	\begin{eqnarray}
    {\cal H}_\mu(B_c\to(\bar{c}c/\bar{u}c)_{S=1})=&&{\frac{1}{(M+m)}}\epsilon ^{\sigma ^\dagger}\Big\{g_{\mu\sigma}(p+k)q A_0(q^2)\nonumber\\
		&&+(p+k)_\mu(p+k)_\sigma A_+(q^2)+q_\mu(p+k)_\sigma A_-(q^2)\nonumber\\&&+i\epsilon_{\mu\sigma\alpha\beta}(p+k)^\alpha q^\beta V(q^2)\Big\} 
	\end{eqnarray}
	The angular decay distribution differential in the momentum transfer squared $q^2$ is obtained in the form \cite{A24} 
	\begin{equation}
	\frac{d\Gamma}{dq^2dcos\theta}	={\frac{{\cal G}_F}{(2\pi)^3}}|V_{bq}|^2\frac{(q^2-m_l^2)^2}{8M^2 q^2}|\vec{k}|{\cal L}^{\mu \sigma}{\cal H}_{\mu \sigma}
	\end{equation}
	Here ${\cal L}^{\mu \sigma}$ and ${\cal H}_{\mu \sigma}$ are the lepton and hadron tensor, respectively; $m_l$ is the mass of charged lepton. The lepton tensor ${\cal L}^{\mu \sigma}=\sum_{\delta_l \delta_\nu}{\cal L}^\mu {\cal L}^{\sigma \dagger}$ is summed over lepton spin indices: $\delta_l,\delta_\nu$. The hadron tensor ${\cal H}_{\mu \sigma}=\sum_{\lambda}{\cal H}_{\mu}(p,k){\cal H}_{\sigma}^{\dagger}(p,k)$ on the other hand, is summed over the daughter meson polarization index $\lambda$.\\
	In our normalization, ${\cal L}^{\mu \sigma}$ is obtained in the form 
	\begin{equation}
		{\cal L}^{\mu \sigma}_{\mp}=8\Bigg\{k^{\mu}_l k^{\sigma}_{\nu}+k^{\sigma}_l k^{\mu}_{\nu}-g^{\mu \sigma}\Bigg({\frac{q^2-m^2_l}{2}}\Bigg)\pm i{\epsilon^{\mu \sigma\alpha \beta}}k_{l_\alpha} k_{\nu_\beta}\Bigg\}
	\end{equation}
	
	\noindent where ${\cal L}_-^{\mu \sigma}$ and ${\cal L}_+^{\mu \sigma}$ refer to the lepton pair: $l\bar{\nu_l}$ and $\bar{l}\nu_l$, respectively. They differ in sign of the parity-odd ${\epsilon}$-tensor contribution. The hadron tensor ${\cal H}_{\mu \sigma}=\sum_{\lambda}{\cal H}_{\mu}(p,k){\cal H}_{\sigma}^{\dagger}(p,k)$ on the other hand, is summed over the daughter meson polarization index $\lambda$. It corresponds to the tensor product of hadronic matrix elements defined above.\\
	
	  It is convenient to express physical observables on a helicity basis. On this basis, the helicity form factors can be expressed  in terms of the Lorentz invariant form factors that represent the decay amplitudes. Then one can perform the Lorentz contraction in Eq(5) with the helicity amplitudes as done in \cite{A24,A51,A52,A53}. For this we consider appropriate helicity projections $\epsilon^\mu(m)$ of the covariants in Eq(3) and (4). There are four covariant helicity projections out of which three projections are orthogonal to the momentum transfer $q$ i.e $\epsilon^\mu(m) q_{\mu}=0$ for $m=\pm,0$ and this constitutes spin 1 part of the $W_{off-shell}$ involved in the decay process. The spin 0 (time-)component m=t of the $W_{off-shell}$ has the property $\epsilon^\mu(t)=\frac{q^\mu}{\sqrt{q^2}}$. The orthogonality and completeness relations satisfied by the helicity projections are
	 
	 	\begin{eqnarray}
	 	\epsilon^{\dagger}_\mu (m)\epsilon^\mu(n)=&&g_{mn} \ \  \ \ \ \     (m,n=t,\pm,0)\nonumber \\
	 	\epsilon_\mu (m)\epsilon^{\dagger}_\sigma(n)g_{mn}=&&g_{\mu \sigma}
	 	 \end{eqnarray} 
	 
	\noindent with $g_{mn}=diag(+,-,-,-)$. Since we would like to study the lepton mass effect in the present investigation of s.l. $B_c$-meson decays, we include the time component polarization $\epsilon^{\mu}(t)$ in addition to its other three components: $m=\pm,0$.\\
	Using the completeness property the lepton and hadron tensors in Eq(5) can be re-written as follows.
	 
	 	\begin{eqnarray}
	 	{\cal L}^{\mu \sigma}{\cal H}_{\mu\sigma}               =&&{\cal L}_{\mu'\sigma'}g^{\mu'\mu}g^{\sigma'\sigma}{\cal H}_{\mu\sigma}\nonumber\\
	 	=&&{{\cal L}_{\mu'\sigma'}}{\epsilon^{\mu^{'}}}(m){\epsilon^{\mu^{\dagger}}}(m^{'}){g_{mm^{'}}}{\epsilon^{\sigma^{\dagger}}}(n){\epsilon^{\sigma^{'}}}(n^{'})g_{nn^{'}}{\cal H}_{\mu\sigma}\nonumber\nonumber\\
	 	=&&{L(m,n)}{g_{mm'}}{g_{nn'}}H({m'}{n'})
	 \end{eqnarray}
	 \\
	 Here lepton and hadron tensors are introduced in the space of helicity components:
	 \begin{eqnarray}
	 	L(m,n)={\epsilon^{\mu}}(m){\epsilon^{\sigma ^\dagger}}(n){\cal L}_{\mu \nu}\nonumber\\
	 	H(m,n)={\epsilon^{\mu^\dagger}}(m){\epsilon^{\sigma}(n)}{\cal H}_{\mu \nu}
	  \end{eqnarray}
  For the sake of convenience, we consider here two frames of reference: i) the $\bar{l}\nu$ or $l\bar{\nu}$ center-of-mass frame and ii)parent $(B_c)$-meson rest frame. We evaluate here the lepton tensor $L(m,n)$ in the  $\bar{l}\nu$ or $l\bar{\nu}$ c.m. frame and hadron tensor $H(m,n)$ in the $B_c$-rest frame. 
  \subsection{Hadron Tensor}
 
  In the $B_c$-rest frame:
\begin{eqnarray}
	p^{\mu}=&&(M,0,0,0)\nonumber\\
	k^{\mu}=&&({E_k},0,0,{-|\vec{k}|})\\
	q^{\mu}=&&({q^0 },0,0,{\vert\vec{k}\vert})\nonumber
\end{eqnarray}

where
	\begin{eqnarray}
	{E_k}=&&\frac{{M^2}+{m^2}-{q^2}}{2M}\nonumber\\
	q^0=&&\frac{{M^2}-{m^2}+{q^2}}{2M}\nonumber\\
	 {E_k}+{q^0}=&&M\\
	 q^{0^2}=&&{q^2}+\vert\vec{k}\vert^2\nonumber\\
	 \vert\vec{k}\vert^2}+{E_k}{{q_0}=&&\frac{1}{2}({M^2}-{m^2}-{q^2})\nonumber
 \end{eqnarray} 
In the $B_c$- rest frame, the polarization vectors of the effective current are:

  \begin{eqnarray}
  	{\epsilon^\mu}(t)=&&\frac{1}{\sqrt{q^2}}{({q_0},0,0,{\vert\vec{k}\vert})}\nonumber\\
  	{\epsilon^\mu}(\pm)=&&\frac{1}{\sqrt{2}}(0,{\mp 1},-i,0)\\
  	{\epsilon^\mu}(0)=&&\frac{1}{\sqrt{q^2}}{({\vert\vec{k}\vert},0,0,{q_0})}\nonumber
  \end{eqnarray}
	
	In the basis (12), the helicity components of the hadronic tensors can be expressed through the invariant form factors defined in Eqs (3) and (4).\\
	For  $B_c\to (\bar{c}c/\bar{u}c)_{S=0}$ transition:
	\begin{equation}
		H(m,n)=\big({\epsilon^{\mu^\dagger}}(m){\cal H}_\mu\big)\big({\epsilon^{\sigma^\dagger}}(m){\cal H}_\sigma\big)^{\dagger}={H_m}{H_n^\dagger}
	\end{equation}
	Then the helicity form factors are expressed in terms of the invariant form factors as  
	
	\begin{eqnarray}
		{H_t}=&&\frac{1}{\sqrt{q^2}}\biggl\{(p+k).(p-k){F_+}+{q^2}{F_-}\biggr\}\nonumber\\
		{H_\pm}=&&0\\
		{H_0}=&&\frac{2M\vert\vec{k}\vert}{\sqrt{q^2}}{F_+}\nonumber
	\end{eqnarray}
 For $B_c\to(\bar{c}c/\bar{u}c)_{S=1}$ transition, the nonvanishing helicity form factors  are given by
 \begin{equation}
 	H_m=\epsilon^{\mu \dagger}(m){\cal H}_{\mu\alpha}\epsilon^{\alpha \dagger}_2(m)\ \ for\ m=\pm,0
 \end{equation}
 	and
 	\begin{equation}
 		H_t=\epsilon^{\mu \dagger}(t){\cal H}_{\mu\alpha}\epsilon^{\alpha \dagger}_2(0)
 	\end{equation}
As in Eq(13), the hadronic tensor, in this case, is also given by $H(m,n)=H_m H_n^{\dagger}$\\
To express the helicity form factors in terms of the invariant form factors (4), it is necessary to specify the helicity components $\epsilon_2(m)\ (m=\pm,0)$ of the polarization vector of the $(\bar{c}c/\bar{u}c)_{S=1}$state. These components are 
\begin{eqnarray}
	{\epsilon_2^\mu}(\pm)=\frac{1}{\sqrt{2}}(0,\pm1,-i,0)\nonumber\\
	\\
	{\epsilon_2^\mu}(0)=\frac{1}{m}({\vert\vec{k}\vert},0,0,{-E_k})\nonumber
\end{eqnarray}
 They satisfy the orthonormality and completeness relations
 
  \begin{eqnarray}
 	{\epsilon_2^{\mu \dagger}}(r){\epsilon_{2\mu}}(s)=-{\delta_{rs}}\nonumber\\
 	\\
 	{\epsilon_{2\mu}}(r){\epsilon_{2\sigma}^\dagger}(s){\delta_{rs}}=-{g_{\mu\nu}}+\frac{{k_\mu}{k_\sigma}}{m^2}\nonumber 	
 \end{eqnarray}
With this specification of the helicity components, the desired relations between the helicity form factors and the invariant form factors are obtained in the form:

\begin{eqnarray}
	{H_t}=&&{\epsilon^{\mu\dagger}}(t){\epsilon^{\alpha\dagger}_2}(0){\cal H}_{\mu \alpha}\nonumber\\
	=&&\frac{1}{(M+m)}\frac{M{\vert \vec{k}\vert}}{m\sqrt{q^2}}\{(p+k) {.} q\ ({-A_0}+{A_+})+{{q^2}{A_-}}\}\nonumber\\
	{H_\pm}=&&{\epsilon^{\mu\dagger}}(\pm){\epsilon^{\alpha\dagger}_2}(\pm){\cal H}_{\mu \alpha}\\
	=&&\frac{1}{(M+m)}\big\{-(p+k) {.}q\ {A_0}\mp 2M\vert\vec{k}\vert V\big\}\nonumber\\
	{H_0}=&&{\epsilon^{\mu\dagger}}(0){\epsilon^{\alpha\dagger}_2}(0){\cal H}_{\mu \alpha}\nonumber\\
	=&&\frac{1}{(M+m)}\frac{1}{2m\sqrt{q^2}}\Big\{-(p+k){.} q ({M^2}-{m^2}-{q^2}){A_0}+4{M^2}\vert\vec{k}\vert^2 {A_+}\Big\} \nonumber
\end{eqnarray}
\subsection{Lepton Tensor}

In the $(l\bar{\nu})$-c.m. frame $({\vec{k}}_l+{\vec{k}}_\nu=\vec{q}=0)$: the relevant four momenta are :
\begin{eqnarray}
{q^\mu}=&&({\sqrt{q^2}},0,0,0)\nonumber\\
{k_\nu^\mu}=&&\Big(\vert\vec{k_l}\vert,{\vert \vec{k_l}\vert \sin\theta\cos\chi},{\vert \vec{k_l}\vert \sin\theta\sin\chi},{\vert \vec{k_l}\vert \cos\theta}\Big)\\
{k_l^\mu}=&&\Big({E_l},{-\vert \vec{k_l}\vert \sin\theta\cos\chi},{-\vert \vec{k_l}\vert \sin\theta\sin\chi},{-\vert \vec{k_l}\vert \cos \theta}\Big)\nonumber
\end{eqnarray}
where ${E_l}=\frac{{q^2}+{m^2_l}}{2{\sqrt{q^2}}}$, ${\vert\vec{k_l}\vert}=\frac{{q^2}-{m_l^2}}{2{\sqrt{q^2}}}$ and decay angles $(\theta,\chi)$ are respectively, the polar and azimuthal angle of the lepton momentum in $(l\bar{\nu})$ c.m. frame. In this frame the longitudinal and time-component polarization vectors are given by
\begin{eqnarray}
	{\epsilon^\mu}(t)=&&\frac{q^\mu}{\sqrt{q^2}}(1,0,0,0)\nonumber\\
	{\epsilon^\mu}(\pm)=&&\frac{1}{\sqrt{2}}(0,\mp,-i,0)\\
	{\epsilon^\mu}(0)=&&(0,0,0,1)\nonumber
\end{eqnarray}
Using Eq.(6) and (21), it is straightforward to evaluate the helicity representation $L(m,n)$ of the lepton tensor.\\

In the present analysis we do not consider the azimuthal $\chi$ distribution of the lepton pair and therefore integrate over the azimuthal angle dependence of the lepton tensor; which yields the differential $(q^2,\cos\theta)$ distribution in the form:
 \begin{eqnarray}
 	\frac{d\Gamma}{d{q^2}\cos\theta}=&&\frac{3}{8}(1+\cos^2\theta)\frac{d\Gamma_U}{dq^2}+\frac{3}{4}\sin^2\theta.\frac{d\Gamma_L}{dq^2}\mp \frac{3}{4} \cos\theta \frac{d\Gamma_P}{dq^2}+\nonumber\\
 	&&\frac{3}{4}\sin^2\theta\frac{d{\tilde{\Gamma}_U}}{dq^2}+\frac{3}{2}\cos^2\theta\frac{d{\tilde{\Gamma}_L}}{dq^2}+\frac{1}{2}\frac{d{\tilde{\Gamma}}_S}{dq^2}+3\cos\theta\frac{d{\tilde{\Gamma}}_{SL}}{dq^2}
 \end{eqnarray} 
The upper and lower signs associated with the parity-violating term in Eq(22) refer to two cases: $l^-\bar{\nu}$ and $l^+{\nu}$, respectively. Out of seven terms in the r.h.s of Eq(22), four terms identified as "tilde" rates $\tilde{\Gamma}_i$ are linked with the lepton mass and other these are lepton mass-independent terms, identified as $\Gamma_i$ and both are related via a flip factor $\frac{m_{l}^2}{2q^2}$ as:
\begin{equation}
	\frac{d{\tilde{\Gamma}}_i}{dq^2}=\frac{m_l^2}{2q^2}\frac{d{\Gamma_i}}{dq^2}
\end{equation}
The tilde rates do not contribute to the vanishing lepton mass limit. They can be neglected for $e$ and $\mu$ modes but they are expected to yield a sizeable contribution to the $\tau$-modes. Therefore the tilde rates are crucial in evaluating the lepton mass effects in the s.l. decay modes. The differential partial helicity rates $\frac{d\Gamma_i}{dq^2}$ are defined by
\begin{equation}
\frac{d\Gamma_i}{dq^2}=\frac{{\cal G}_f^2}{(2\pi)^3}{\vert V_{bq^{'}}\vert}^2 \frac{({q^2}-{m_e^2})^2}{12{m_1^2}q^2}|\vec{k}|H_i
\end{equation}

Here $H_i(i=U,L,P,S,SL)$ represents a standard set of helicity structure function given by linear combinations of helicity components of hadron tensor $H(m,n)=H_m H_n^\dagger$:

\begin{eqnarray}
	{H_U}=&&Re({H_+}{H_+^\dagger})+Re({H_-}{H_-^\dagger})\ \ \ \ \   :Unpolarized-transversed\nonumber\\
	{H_L}=&& Re({H_0}{H_0^\dagger})\ \ \ \ \ \ \ \ \ \ \ \ \ \ \ \ \ \ \ \ \ \ \ \  :Longitudinal\nonumber\\
	{H_P}=&&Re({H_+}{H_+^\dagger})-Re({H_-}{H_-^\dagger})\ \ \ \ :Parity-odd\nonumber\\
	{H_S}=&&3Re({H_t}{H_t^\dagger})\ \ \ \ \ \ \ \ \ \ \ \ \ \ \ \ \ \ \ \ \ \ :Scalar\nonumber\\
	{H_{SL}}=&&Re({H_t}{H_0^\dagger})\ 
	\ \ \ \ \ \ \ \ \ \ \ \ \ \ \ \ \ \ \ \ \ \ :Scalar-Longitudinal\ Interference\nonumber
\end{eqnarray}
Here we assume throughout that the helicity amplitudes are real since the available $q^2$- range: ($q^2\le(M-m)^2$) is below the physical threshold $q^2= (M+m)^2$. Therefore we drop the angular terms that are multiplied by coefficients $Im(H_iH_j^*),i\ne j^*$.\\

Then integrating over $\cos\theta$ one gets the differential $q^2$ distribution and finally integrating over $q^2$, one obtains the total decay rate $\Gamma$ as the sum of the partial decay rates :

$\Gamma_i=(i=U,L,P)$ and $\tilde{\Gamma}_i(i=U,L,S,SL)$.\\

A quantity of interest is the forward-backward asymmetry $A_{FB}$ of the lepton in the $(l\bar{\nu})$ c.m. frame which is defined as\\
\begin{equation}
A_{FB}=\frac{3}{4}\biggl\{\frac{\pm P+4\tilde{SL}}{U+\tilde{U}+L+\tilde{L}+\tilde{S}}\biggr\}
\end{equation}
Another quantity of interest is the asymmetry parameter $\alpha^*$ which is defined by rewriting  Eq(22) in terms of its $\cos^2\theta^*$ dependence i.e $d\Gamma \propto 1+\alpha^* \cos ^2\theta^*$. The asymmetry parameter $\alpha^*$ which determines the transverse and longitudinal composition of the vector meson, the final state is given by: 
\begin{eqnarray}
	\alpha^*=\frac{U+\tilde{U}-2(L+\tilde{L}+\tilde{S})}{U+\tilde{U}+2(L+\tilde{L}+\tilde{S})} 
\end{eqnarray}
We list our predictions on helicity rates $\Gamma_i$, $\tilde{\Gamma}_i$, $A_{FB}$ and $\alpha^*$ in Sec. IV.
\section{Transition Matrix element and weak form factors}
The decay process physically takes place when participating mesons are in their momentum eigenstates. Therefore, in the field-theoretic description of any decay process, it is necessary to represent the meson bound-states by appropriate momentum wave-packets reflecting momentum and spin distribution between constituent quark and antiquark inside the meson core. In the RIQ model, the wavepacket representing a meson bound state $\vert B_c(\vec{p}, S_{B_c})\rangle$, for example, at a definite momentum $\vec{p}$ and spin $S_{B_c}$ is taken in the form\cite{A29,A30,A31,A32,A33,A34,A35}
\begin{equation}
	\vert B_c(\vec{p},S_{B_c})\rangle=\hat{\Lambda}(\vec{p},S_{B_c})\vert (\vec{p_b},\lambda_b);(\vec{p_c},\lambda_c)\rangle 
\end{equation}
where $\vert (\vec{p_b},\lambda_b);(\vec{p_c},\lambda_c)\rangle $ is the fock space representation of the unbound quark and antiquark in a color-singlet configuration with their respective momentum and spin: $(\vec{p_b},\lambda_b)$ and $(\vec{p_c},\lambda_c)$. Here $\hat{b_b}^\dagger(\vec{p_b},\lambda_b)$  and $\hat{b_c}^\dagger(\vec{p_c},\lambda_c)$ denote the quark-antiquark creation operator, respectively and $\hat{\Lambda}(\vec{p}, S_{B_c})$ is taken as a bag like integral operator in the form:
\begin{equation}
	\hat{\Lambda}(\vec{p},S_{B_c})=\frac{\sqrt{3}}{\sqrt{N(\vec{p})}}\sum_{\delta_b \delta_c}\zeta_{b_c}^{B_c}(\lambda_b,\lambda_c)\int d^3p_bd^3p_c\delta^{(3)}(\vec{p_b}+\vec{p_c}+\vec{p}){\cal G}_{B_c}(\vec{p_b},\vec{p_c})
\end{equation}
Here $\sqrt{3}$ is the effective color factor,$\zeta_{b_c}^{B_c}(\lambda_b,\lambda_c)$ is the $SU(6)$ spin-flavor coefficients for $B_c$-meson and $N(\vec{p})$ is the meson-state normalization obtained in an integral form:
\begin{equation}
	N(\vec{p})=\int d^3\vec{p_b}\vert {\cal G}_{B_c}(\vec{p_b},\vec{p}-\vec{p_b})\vert^2
\end{equation}
by imposing the normalisation condition $\langle{B_c}(\vec{p})\vert{B_c}(\vec{p}^{'})\rangle=\delta^3 (\vec{p}-\vec{p}^{'})$. Finally ${\cal G}_{B_c}(\vec{p_b},\vec{p}-\vec{p_b})$ is the effective momentum distribution function for the quark(b) and antiquark(c) pair in the meson core. In terms of individual momentum probability amplitudes: $G_b(\vec{p_b})$ and $G_c(\vec{p_c})$ of the constituent quarks, ${\cal G}_{B_c}(\vec{p_b},\vec{p}-\vec{p_b})$ is taken in the form:
\begin{equation}
	{\cal G}_{B_c}(\vec{p_b},\vec{p}-\vec{p_b})=\sqrt{G_b(\vec{p_b})G_c(\vec{p}-\vec{p_b})}
\end{equation}
in the straightforward extension of the ansatz of Margolis and Mendel in their bag model description \cite{A54}. The quark orbitals derived in the framework of the RIQ model and corresponding momentum probability amplitudes are briefly discussed in the 'Appendix'.\\
\begin{figure}
	\includegraphics{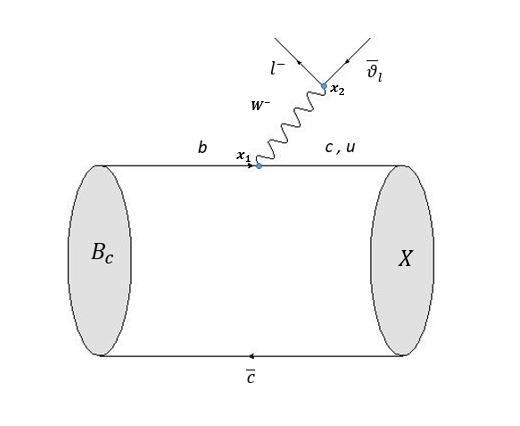}\caption{Semileptonic decay of $B_c$ meson}
	\label{FD}
\end{figure}
In the wavepacket representation of meson bound-states (27-29), the effective momentum distribution function, here, embodies the bound-state character inherent in $\vert B_c(\vec{p}, S_{B_c})\rangle$. Any residual internal dynamics responsible for the decay process can be described at the level of otherwise unbound quark and antiquark using usual the Feynman technique. In the straightforward calculation of the appropriate Feynman diagram shown in FIG.\ref{FD}, the constituent level $S-$matrix element $S_{fi}^{b\to c,u}$ is obtained for the decay process induced by $b\to c,u$ transitions. The quark level $S-$matrix element $S_{fi}^{b\to c,u}$ when operated upon by the bag like operator $\hat{\Lambda}$ in (28) yields the mesonic level $S-$matrix in the form:
\begin{equation}
	S_{fi}^{{B_c}\to(\bar{c}c/\bar{u}c)system}\longrightarrow \hat{\Lambda}S_{fi}^{b\to c/u}
\end{equation}
\subsection{Transition amplitude}
The $S-$matrix element for the decay process $B_c\to Xl\bar{\nu_l}$ depicted in FIG.\ref{FD} is written in general form:
\begin{equation}
	S_{fi}=-i\frac{{\cal G}_F}{\sqrt{2}}V_{bc/u}\frac{1}{(2\pi)^4}\langle l(\vec{k_l},\delta_l)\bar{\nu_l}(\vec{k_\nu},\delta_\nu)X(\vec{k})\vert J_l^\mu (x_2)J_\mu^h(x_1)\vert B_c(\vec{p},S_{B_c})\rangle
\end{equation}
where the leptonic weak current contribution is 
\begin{equation}
	\langle l(\vec{k_l},\delta_l)\bar{\nu_l}(\vec{k_\nu},\delta_\nu)\vert J_l^\mu(x_2)\vert 0\rangle =\frac{e^{i(k_i+k_\nu)x_2}}{(2\pi)^3\sqrt{2E_l2E_\nu}}{\cal L}^\mu  
\end{equation} 
 with 
 \begin{equation}
 	{\cal L}^\mu=\sum_{\delta_l\delta_\nu}\bar{u}(k_l,\delta_l)\Gamma^\mu v(k_\nu,\delta_\nu),\ \ \ \ \  \Gamma ^\mu =\gamma^\mu(1-\gamma_5) 
 \end{equation}	  
 This along with the hadronic amplitude $'{\cal H}'$ obtained from the overlapping integral of meson wavefunctions in terms of the wave packet representation of the participating  meson states, one gets the $S-$matrix element for the decay process in the standard form:
 \begin{equation}
 	S_{fi}=(2\pi)^4\delta^{(4)}(p-k-k_l-k_\nu)(-{\cal M}_{fi})\frac{1}{\sqrt{(2\pi)^32E_{B_c}}}\Pi_f \frac{1}{\sqrt{2E_f(2\pi)^3}} 
 \end{equation} 
The hadronic amplitude ${\cal H}_\mu$ in the $B_c-$ rest frame is obtained as:
\begin{equation}
	{\cal H}_\mu=\sqrt{\frac{4ME_k}{N_{B_c}(0)N_X(\vec{k})}}\int\frac{d^3p_b}{\sqrt{2E_{p_b}2E_{k+p_b}}}{\cal G}_{B_c}(\vec{p_b},-\vec{p_b}){\cal G}_X(\vec{k}+\vec{p_b},-\vec{p_b})\langle S_X\vert J_\mu^h(0)\vert S_{B_c}\rangle  
\end{equation} 	  
 	  where $E_{p_b}$ and $E_{p_b+k}$ stand for the energy of the non-spectator quark of the parent and daughter meson, respectively, and $\langle S_X\vert J_\mu^h(0)\vert S_{B_c}\rangle $ represents symbolically the spin matrix elements of vector-axial vector current.
 	  \subsection{Weak decay form factors}
For $0^-\to 0^-$ transitions, the axial vector current does not contribute. The spin matrix elements corresponding to the non-vanishing vector current parts are obtained in the form:
\begin{eqnarray}
	\langle S_X(\vec{k})\vert V_0\vert S_{B_c}(0)\rangle =\frac{(E_{p_b}+m_b)(E_{p_{c/u}}+m_{c/u})+|\vec{p_b}|^2}{\sqrt{(E_{p_b}+m_b)(E_{p_{c/u}}+m_{c/u})}}\\
	\langle S_X(\vec{k})\vert V_i\vert S_{B_c}(0)\rangle =\frac{(E_{p_b}+m_b)k_i}{\sqrt{(E_{p_b}+m_b)(E_{p_b+k}+m_{c/u})}}
\end{eqnarray}	  
 	  With the above spin matrix elements, the expressions for hadronic amplitudes (36) are compared with corresponding expressions  in Eq(3) yielding the form factors $f_+$ and $f_-$ for $0^-\to 0^-$ transition in the form :
 	  \begin{eqnarray}
 	  f_\pm (q^2)=&&\frac{1}{2M}\sqrt{\frac{ME_k}{N_{B_c}(0)N_X(\vec{k})}}\int d\vec{p_b}{\cal G}_{B_c}(\vec{p_b},-\vec{p_b}){\cal G}_X(\vec{k}+\vec{p_b},-\vec{p_b})\nonumber\\ && \times \frac{(E_{o_b}+m_b)(E_{p_{c/u}}+m_{c/u})+|\vec{p_b}|^2\pm(E_{p_b}+m_b)(M\mp E_k)}{E_{p_b}E_{p_{c/u}}(E_{p_b}+m_b)(E_{p_{c/u}}+m_{c/u})}	
 	  \end{eqnarray}
 	  For $(0^-\to 1^-)$ transitions, the spin matrix elements corresponding to the vector and axial-vector current are found separately in the form:
 	  \begin{eqnarray}
 	  \langle S_X(\vec{k},\hat{\epsilon^*})\vert V_0\vert S_{B_c}(0)\rangle =&&0\\
 	  \langle S_X(\vec{k},\hat{\epsilon^*})\vert V_i\vert S_{B_c}(0)\rangle=&&\frac{i(E_{p_b}+m_b)(\hat{\epsilon}^*\times \vec{k})_i}{\sqrt{(E_{p_b}+m_b)(E_{p_b+k}+m_{c/u})}}\\
 	  \langle S_X(\vec{k},\hat{\epsilon^*})\vert A_i\vert S_{B_c}(0)\rangle=&&\frac{(E_{p_b}+m_b)(E_{p_b+k}+m_{c/u})-\frac{|\vec{p_b}|^2}{3}}{\sqrt{(E_{p_b}+m_b)(E_{p_b+k}+m_{c/u})}}\\
 	  \langle S_X(\vec{k},\hat{\epsilon^*})\vert A_0\vert S_{B_c}(0)\rangle=&&\frac{-(E_{p_b}+m_b)(\hat{\epsilon}^*. \vec{k})}{\sqrt{(E_{p_b}+m_b)(E_{p_b+k}+m_{c/u})}}
 	  \end{eqnarray}
 	  
 	 With the spin matrix elements (40-43), the expressions for hadronic amplitudes (36) are compared with corresponding expressions in Eq(4). The model expressions for form factors: $V(q^2),A_0(q^2),A_+(q^2)$ and $A_-(q^2)$ are obtained in the form:
 	 \begin{eqnarray}
 	 	V(q^2)=&&\frac{M+m}{2M}\sqrt{\frac{ME_k}{N_{B_c}(0)N_X(\vec{k})}}\int d\vec{p_b}{\cal G}_{B_c}(\vec{p_b},-\vec{p_b}){\cal G}_X(\vec{k}+\vec{p_b},-\vec{p_b})\nonumber\\ &&\times \sqrt{\frac{(E_{p_b}+m_b)}{E_{p_b}E_{p_{c/u}}(E_{p_{c/u}}+m_{c/u})}}	\\
 	 	A_0(q^2)=&&\frac{1}{(M-m)}\sqrt{\frac{Mm}{N_{B_c}(0)N_X(\vec{k})}}\int d\vec{p_b}{\cal G}_{B_c}(\vec{p_b},-\vec{p_b}){\cal G}_X(\vec{k}+\vec{p_b},-\vec{p_b})\nonumber\\ &&\times \frac{(E_{p_b}+m_b)(E^0_{p_{c/u}}+m_{c/u})-\frac{|\vec{p_b}|^2}{3}}{\sqrt{E_{p_b}E_{p_{c/u}}(E_{p_b}+m_b)(E_{p_{c/u}}+m_{c/u})}}
  	\end{eqnarray}
  with $E^0_{p_{c/u}}=\sqrt{|\vec{p}_{c/u}|^2+m^2_{c/u}}$
  and\begin{equation}
  	A_\pm(q^2)=\frac{-E_k(M+m)}{2M(M+2E_k)}\big[T\mp \frac{3(M\mp E_k)}{(E_k^2-m^2)}\big\{I-A_0(M-m)\big\}\big]
  \end{equation}
where $T=J-(\frac{M-m}{E_k})A_0$,\\
with \begin{eqnarray}
	J=&&\sqrt{\frac{ME_k}{N_{B_c}(0)N_X(\vec{k})}}\int d\vec{p_b}{\cal G}_{B_c}(\vec{p_b},-\vec{p_b}){\cal G}_X(\vec{k}+\vec{p_b},-\vec{p_b})\nonumber\\
	&&\times \sqrt{\frac{(E_{p_b}+m_b)}{E_{p_b}E_{p_{c/u}}(E_{p_{c/u}}+m_{c/u})}}\\
	I=&&\sqrt{\frac{ME_k}{N_{B_c}(0)N_X(\vec{k})}}\int d\vec{p_b}{\cal G}_{B_c}(\vec{p_b},-\vec{p_b}){\cal G}_X(\vec{k}+\vec{p_b},-\vec{p_b})\nonumber\\
	&&\times \biggl\{\frac{(E_{p_b}+m_b)(E^0_{p_{c/u}}+m_{c/u})-\frac{|\vec{p}_b|^2}{3}}{\sqrt{E_{p_b}E^0_{p_{c/u}}(E_{p_b}+m_b)(E^0_{p_{c/u}}+m_{c/u})}}\biggr\}
\end{eqnarray}
With the relevant form factors thus obtained in terms of model quantities, the helicity amplitudes and hence the decay rates for $B_c\to \eta_c(J/\psi)l\bar{\nu}_l$ and $B_c\to D(D^*)l\bar{\nu}_l$ are evaluated and our predictions are listed in the next section.
\section{Numerical results and Discussion}
\begin{figure}[hbt!]
\includegraphics[width=.4\textwidth]{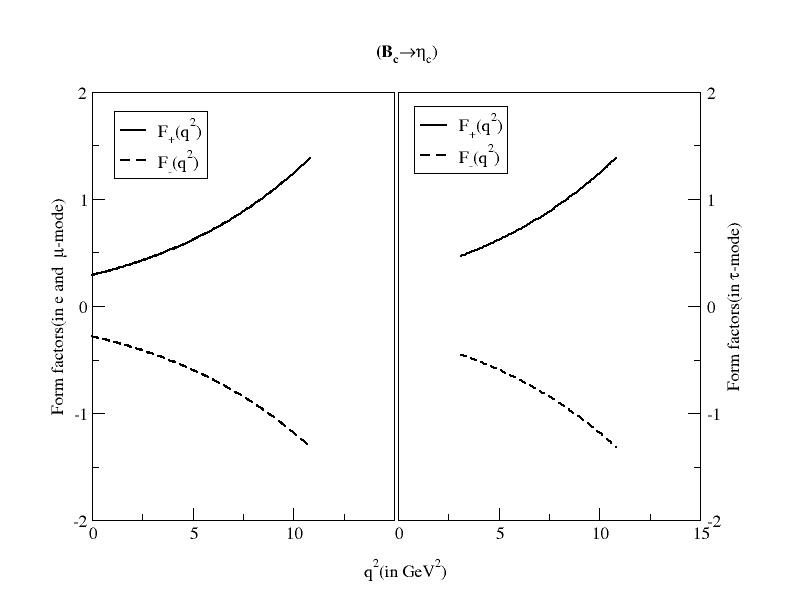}
\includegraphics[width=.4\textwidth]{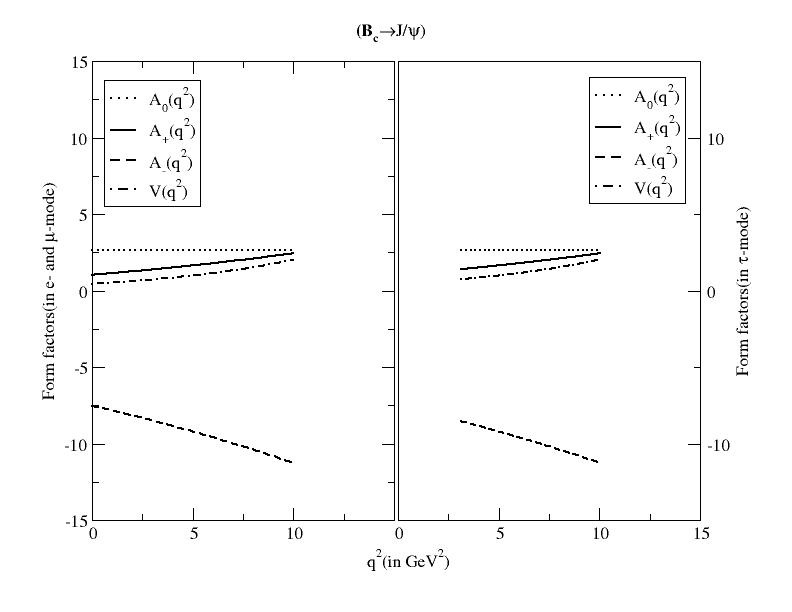}

	\caption{The $q^2$-dependence of invariant form factors for semileptonic $B_c\to \eta_c (J/\psi)$ decays.}
	\label{ff eta-c J-psi}
\end{figure}

\begin{figure}[hbt!]
\includegraphics[width=.4\textwidth]{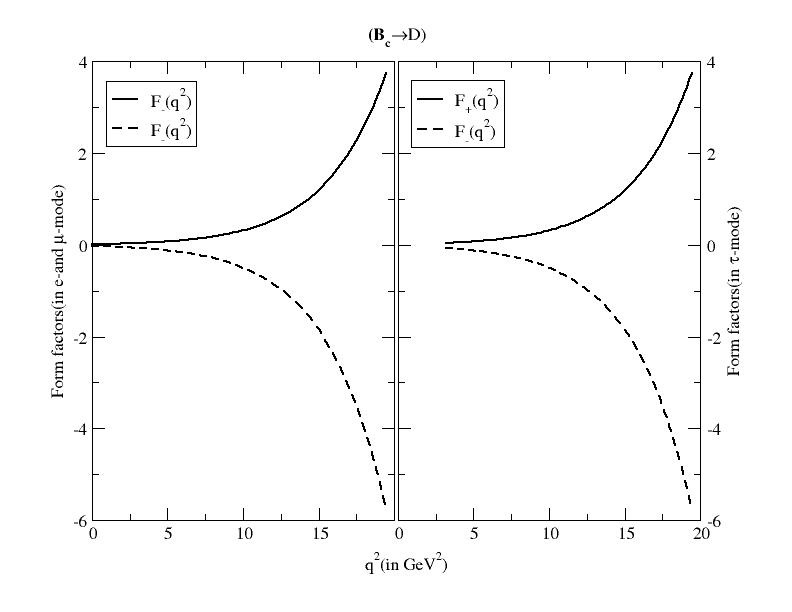}
\includegraphics[width=.4\textwidth]{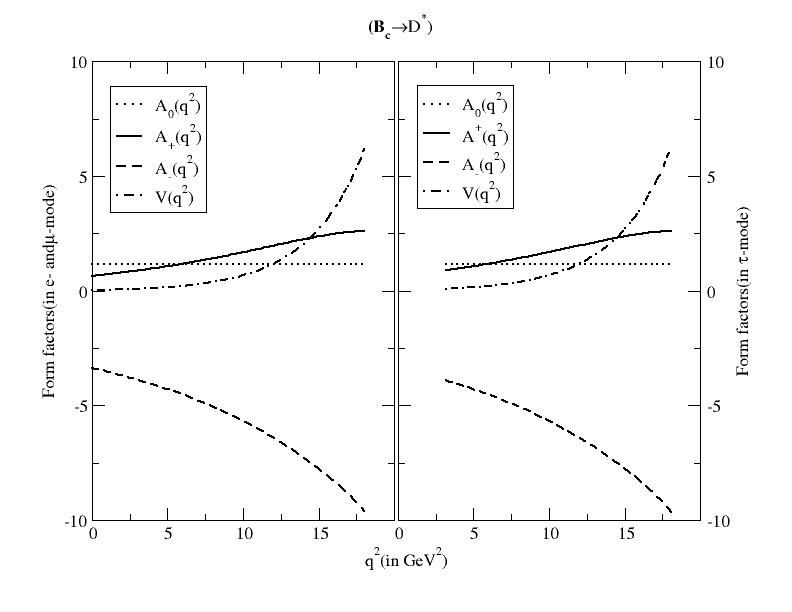}
\caption{The $q^2$-dependence of invariant form factors for semileptonic $B_c\to D(D^*)$ decays.}
\label{Dstarff}
\end{figure}

In this section, we present the numerical results on exclusive semileptonic decays: $B_c\to \eta_c(J/\psi)l\bar{\nu}_l$ and $B_c\to D(D^*)l\bar{\nu}_l$. The dynamics underlying the decay process are well understood in the framework of the suitable phenomenological model by comparing the model predictions on observables with other model predictions. For numerical analysis, we use the RIQ model parameters $(a, V_0)$, quark mass $'m_q'$, and quark binding energy $'E_q'$ as \cite{A28} 
\begin{eqnarray}
	(a, V_0)=&&(0.017166\ GeV^3,-0.1375\ GeV)\nonumber\\
	(m_b,m_c,m_u)=&&(4.77659,1.49276,0.07875)\ GeV\\
	(E_b,E_c,E_u)=&&(4.76633,1.57951,0.47125)\ GeV\nonumber
\end{eqnarray}
 The input parameters (49) have been fixed in the early application of the RIQ model by fitting the data of heavily flavored meson while reproducing hyperfine mass-splitting in the heavy flavor sector. Using the same set of input parameters, wide-ranging hadronic phenomena have been described \cite{A29,A30,A31,A32,A33,A34,A35} in this model. For CKM-parameters and $B_c-$meson lifetime, we take their central values from  the Particle Data Group(2020)\cite{A50} as:
 \begin{eqnarray}
 	(V_{bc},V_{bu})=&&(0.041,0.00382)\nonumber\\
 	\tau _{B_c}=&&0.507\ ps
 \end{eqnarray}   
and for the physical mass of participating mesons, we take corresponding recorded values \cite{A50} as:
\begin{eqnarray}
	M\equiv M_{B_c}=&&6.2749\ GeV\nonumber\\
	(m_{\eta_c},m_{J/\psi})=&&(2.9839,3.0969)\ GeV\\
	(m_D,m_{D^*})=&&(1.86483,2.0068)\ GeV\nonumber
\end{eqnarray}
With these input parameters (49-51) the Lorentz invariant form factors: $(F_+, F_-; A_0, A_+ ,A_-, V)$ representing decay amplitudes can be calculated from the overlapping integral of participating meson wave functions. We first study the $q^2-$dependence of the invariant form factors in the allowed kinematic range. We  plot it in FIG. \ref{ff eta-c J-psi} and \ref{Dstarff} for s.l. $B_c$-decays: $B_c\to \eta_c (J/\psi)$ and $B_c\to D(D^*)$, respectively in their $e^-,\mu^-$ and $\tau^-$-modes over the accessible kinematic range. We find the behavior of form factors in $e^-$ and $\mu ^-$-mode overlap in the entire kinematic range: $0\le q^2\le q^2_{max}$. This is because of an insignificant change in the phase space  boundary going from $e^-$ to $\mu^-$ mode and as one can see here the maximal lepton energy shift $\frac{m^2_\mu-m^2_e}{2M}$ is invisible at the usual scale of the plot. On the other hand for the $\tau^-$ mode  decays, the relevant form factors behave differently throughout the accessible kinematic range of $q^2_{min}\le q^2\le q^2_{max}$, where $q^2_{min}$ is +ve and away from $q^2\to 0$. The $\tau-$phase space, as compared to the $e^-$ and $\mu ^-$ cases is considerably reduced and shifted to a large $q^2-$region. Therefore, in the present analysis, we shall consider decays in the $e^-$ and $\tau ^-$modes only for evaluating the lepton mass effects on the s.l. $B_c$ meson decays.\\

As mentioned earlier it is convenient to calculate decay amplitudes in the helicity basis in which the partial helicity rate and total decay rates are expressed in terms of helicity form factors. On this basis, the expressions for the relevant helicity form factors have also been obtained in terms of invariant form factors (14,19). Using the helicity form factors it is straightforward to obtain partial helicity rates:$\frac{d\Gamma_i}{dq^2}$ (without flip) and $\frac{d\tilde{\Gamma}_i}{dq^2}$(with flip) for all four decay modes considered here. Since our main objective is to evaluate the lepton mass effects on s.l. decays, we would like to study the $q^2$-dependence  of the helicity form factors and partial helicity rates as well as $q^2$-spectra of s.l. decay rates, separately in their $e^-$ and $\tau ^-$ modes. For this, we first rescale the helicity form factors according to

\begin{equation}
	h_j=A(q^2)H_j,\ \ \ \ j=0,+,-\ \ \ (for\  no\ flip\ case)
\end{equation}   
 	and for flip cases
 	\begin{eqnarray}
 		\tilde{h}_j=&&\sqrt{\frac{m_l^2}{2q^2}}A(q^2)H_j\ ,\ \ \ \ \ j=0,+,-\nonumber\\
 		\tilde{h}_t=&&\sqrt{\frac{m_l^2}{2q^2}}\sqrt{3}A(q^2)H_t
 	\end{eqnarray}

 where \begin{equation}
 	A(q^2)=\frac{{\cal G}_F}{4M}\big(\frac{q^2-m_l^2}{q^2}\big)\sqrt{\frac{|\vec{k}|q}{6\pi^3}}\ |V_{b,c/u}|
 \end{equation}
and $\sqrt{\frac{m_l^2}{2q^2}}$: denotes a flip factor.\\
In terms of the rescaled  helicity form factors, the angle integrated differential $q^2$-rate is expressed as:
\begin{equation}
	\frac{d\Gamma}{dq^2}=\sum_{0,+,-}|h_j|^2+\sum_{t,0,+,-}|\tilde{h}_j|^2 
\end{equation}

\begin{figure}[hbt!]
\includegraphics[width=.4\textwidth]{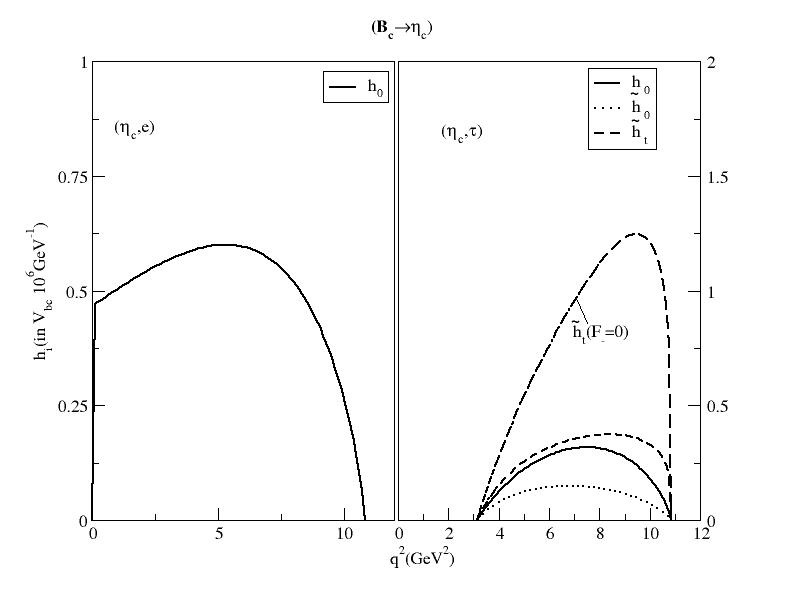}
\includegraphics[width=.4\textwidth]{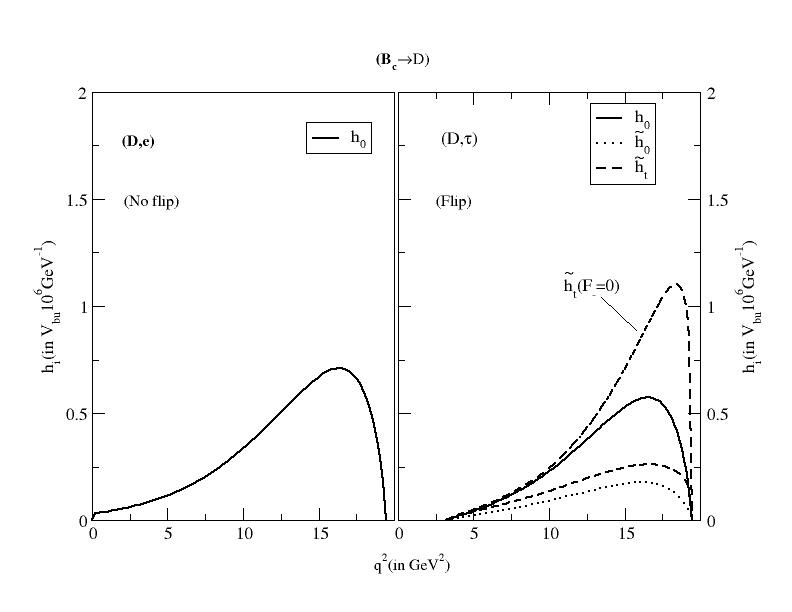}
\caption{Reduced helicity amplitudes $h_i$ and $\tilde{h}_i(i=t,0)$ as functions of $q^2$ for semileptonic $B_c\to \eta_c$ and $B_c\to D$ decays.}
\label{eta-c-D-hr}
\end{figure}
\begin{figure}[hbt!]

\includegraphics[width=.4\textwidth]{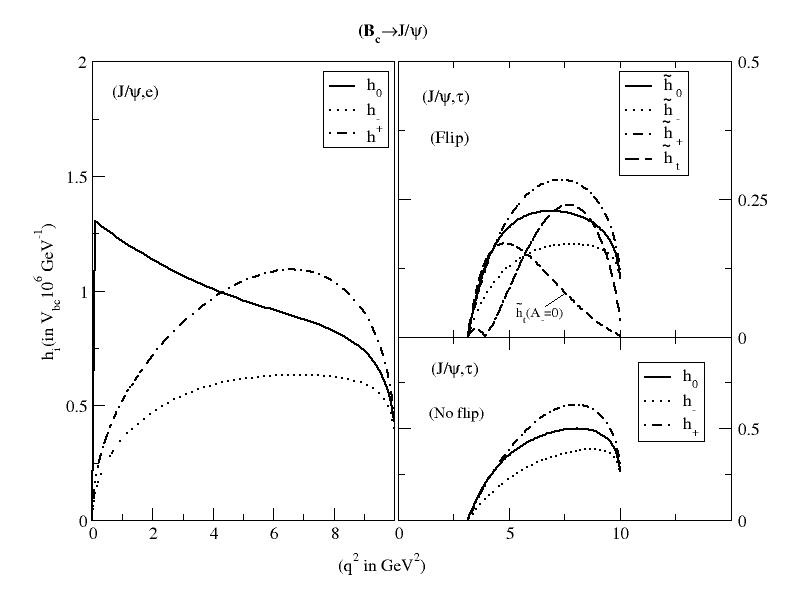}
\includegraphics[width=.4\textwidth]{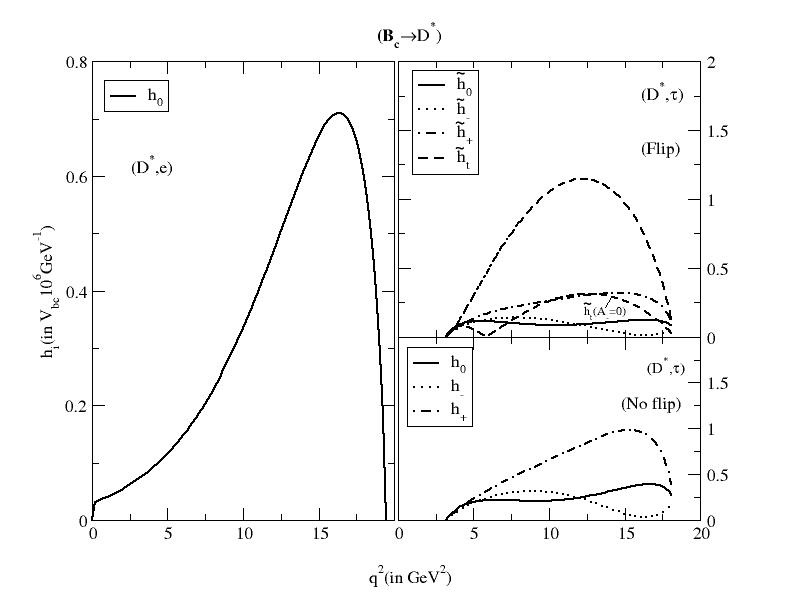}
\caption{Reduced helicity amplitudes $h_i$ and $\tilde{h}_i(i=t,+,-,0)$ as functions of $q^2$ for semileptonic $B_c\to J/\psi$ and $B_c\to D^*$ decays.}
\label{J-psi-D*-hr}

\end{figure}

 	  In FIG.\ref{eta-c-D-hr} we plot the $q^2-$dependence of the rescaled helicity form factors $h_j$ and $\tilde{h}_j$ for $B_c\to \eta_c$ and $B_c\to D$ decays, respectively in their $e^-$ and $\tau^-$modes. We find in both the decay processes, the longitudinal no-flip amplitude $h_0$ is reduced in the low $q^2$ region going from $e^-$ to $\tau^-$ mode. This reduction is due to the threshold-like factor; $\frac{q^2-m^2_l}{q^2}$ appearing in the rescaled helicity amplitude. The longitudinal flip amplitude $\tilde{h}_0$ is further reduced by the flip factor $\sqrt{\frac{m_l^2}{2q^2}}$. The large value of the scalar flip amplitude $\tilde{h}_t$ is attributed to the fact that time-like (scalar-)current contribution here proceeds via an orbital $S-$wave, where there is no pseudo-threshold  factor to tamper the enhancement at large $q^2$ resulting from the time-like form factor in the helicity amplitude.\\
 	  
 	  In FIG.\ref{J-psi-D*-hr} we plot the $q^2$-dependence of the rescaled helicity amplitudes for $B_c\to J/\psi$ and $B_c\to D^*$ transitions in their $e^-$ and $\tau^-$ modes. The largest reduction occurs for longitudinal no-flip amplitude $h_0$. Contrary to $B_c\to \eta_c(D)$ cases, we find here all flip amplitudes generally small compared to no-flip amplitudes. This is attributed to the partial wave structure of the scalar-current contribution. The effects of the time-like(scalar)-current on s.l. decays in $\tau$-mode have been depicted in FIG.\ref{eta-c-D-hr} and \ref{J-psi-D*-hr}. It is interesting to see how the invariant form factors: $F_-(q^2)$ and $A_-(q^2)$ dominate to determine the shape of the plot of flip helicity component $\tilde{h}_t$ over the accessible kinetic range. In $B_c\to\eta_c(D)$ cases, the contribution of $F_-(q^2)$ to the time-like helicity form factor is destructive. Hence, corresponding rescaled helicity amplitude $\tilde{h}_t$ increases when $F_-(q^2)$ is switched off, as shown in FIG.\ref{eta-c-D-hr}. Since $|\tilde{h}_t|^2$ gives dominant contribution to the decay rate, an accurate determination of the s.l. $B_c\to \eta_c$ and $B_c\to D$ decays in their $\tau$-mode would allow one to extract information on the sign and magnitude of the scalar-invariant form factors. On the other hand, the contribution of scalar form factor $A_-(q^2)$ in $B_c\to J/\psi (D^*)$ cases is constructive. Consequently, $\tilde{h}_t$ decreases when $A_-(q^2)$ is switched off. This is shown in FIG.\ref{J-psi-D*-hr}.\\
 	  
 	  \begin{figure}
 	  	\includegraphics[width=.4\textwidth]{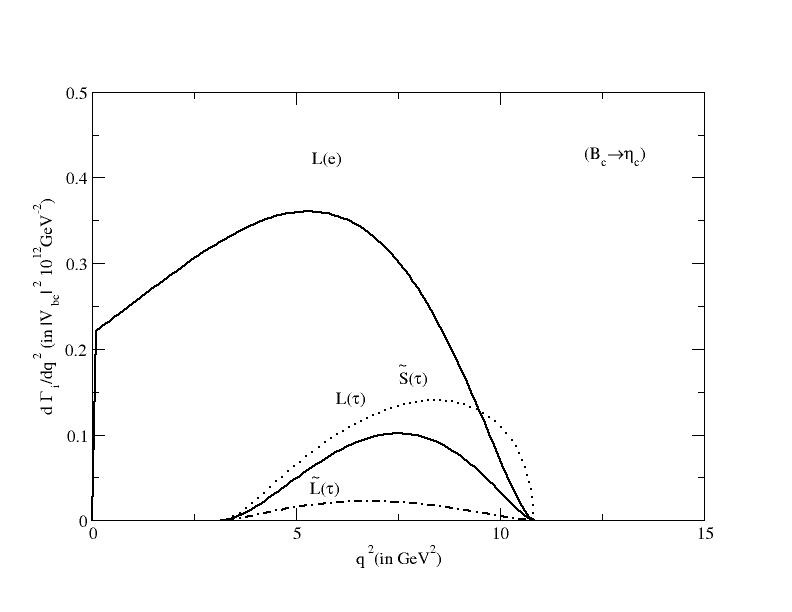}
 	  		\includegraphics[width=.4\textwidth]{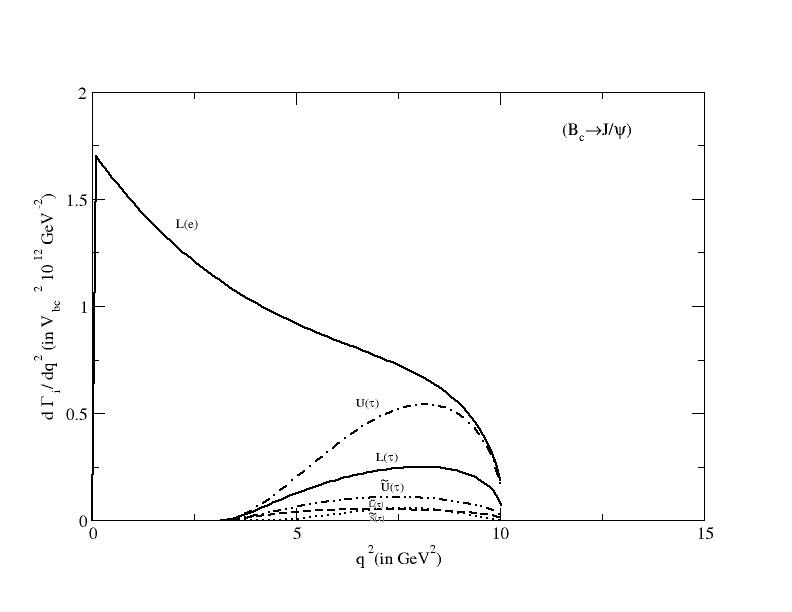}
 	  		\caption{Partial helicity rates $\frac{d\Gamma_i}{dq^2}$ and $\frac{d\tilde{\Gamma}_i}{dq^2}$ as functions of $q^2$ for semileptonic $B_c\to \eta_c$ and $B_c\to D$ decays}
			\label{fig 6}
 	  		\end{figure}
 \begin{figure}
 	\includegraphics[width=.4\textwidth]{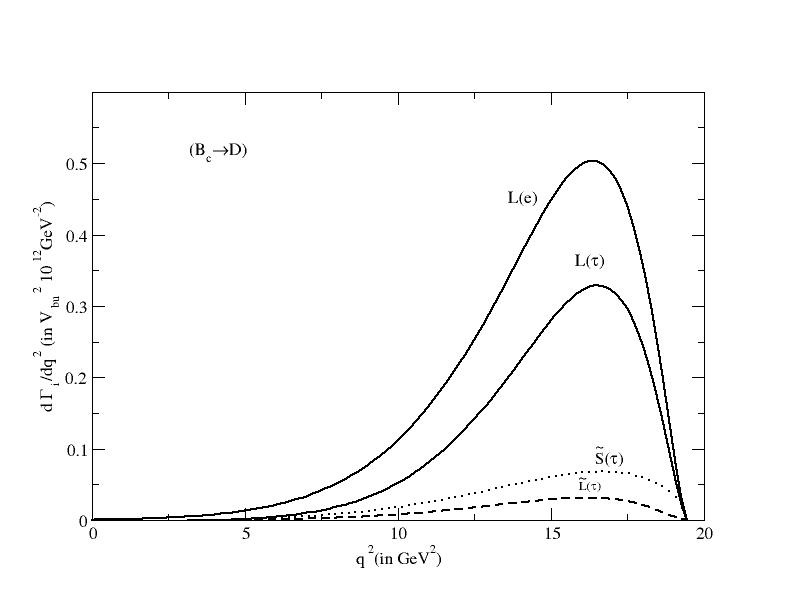}
 	  \includegraphics[width=.4\textwidth]{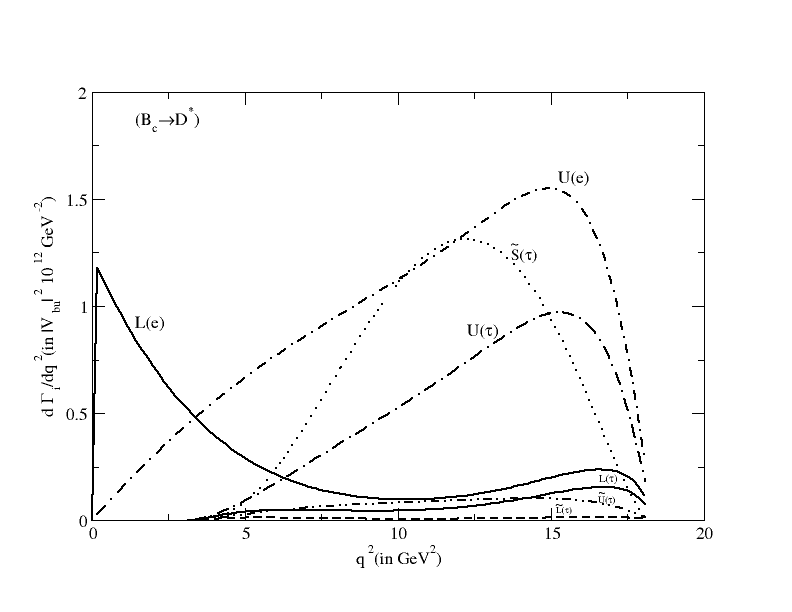}
 	  \caption{Partial helicity rates $\frac{d\Gamma_i}{dq^2}$ and $\frac{d\tilde{\Gamma}_i}{dq^2}$ as functions of $q^2$ for semileptonic $B_c\to J/\psi$ and $B_c\to D^*$ decays}
 	  \label{fig 7}
 	  	
 	  \end{figure}

 	  In FIG.\ref{fig 6} we plot the $q^2$-spectra for $B_c\to \eta_c(e^-,\tau^-)$ and $B_c\to D(e^-,\tau^-)$ decay modes for different helicity contributions. In these cases we find a considerable reduction of longitudinal no-flip contribution, going from $e^-$ and $\tau^-$ mode. It is also noteworthy to find that, a sizeable contribution to the differential decay rate here comes from the scalar-flip component over the allowed kinematic range.\\
 	  
 	  In FIG.\ref{fig 7} we plot the $q^2$-spectra for $B_c\to J/\psi(e^-,\tau^-)$ and $B_c\to D^*(e^-,\tau ^-)$ modes. We find here the spin-flip component negligible compared to no-flip parts except in the $B_c\to D^*$ case where the scalar flip part $\tilde{S}$ of the helicity amplitude provides a sizeable contribution in high $q^2$-region. The helicity rates appear to be uniformly reduced going from $e^-$ to $\tau^-$ mode, except for the longitudinal contribution which is disproportionately reduced due to threshold-like factor $\frac{q^2-m_l^2}{q^2}$.\\
 	  
  \begin{figure}[hbt]

  	\includegraphics[width=.4\textwidth]{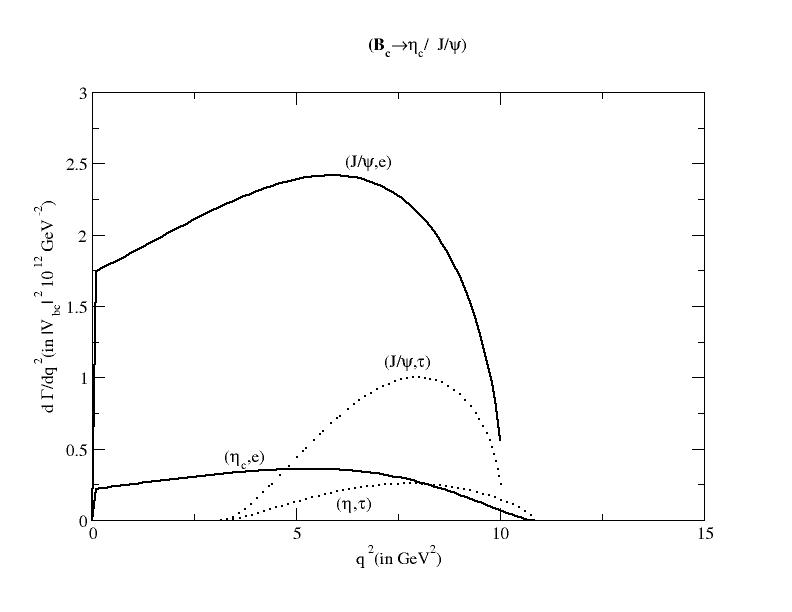}\caption{$q^2$-spectrum of s.l. decay rates for $B_c\to \eta_c$ and $B_c\to J/\psi$ decays.}
  	\label{fig 8}
  \end{figure}
  \begin{figure}[hbt]
  	
  	\includegraphics[width=.4\textwidth]{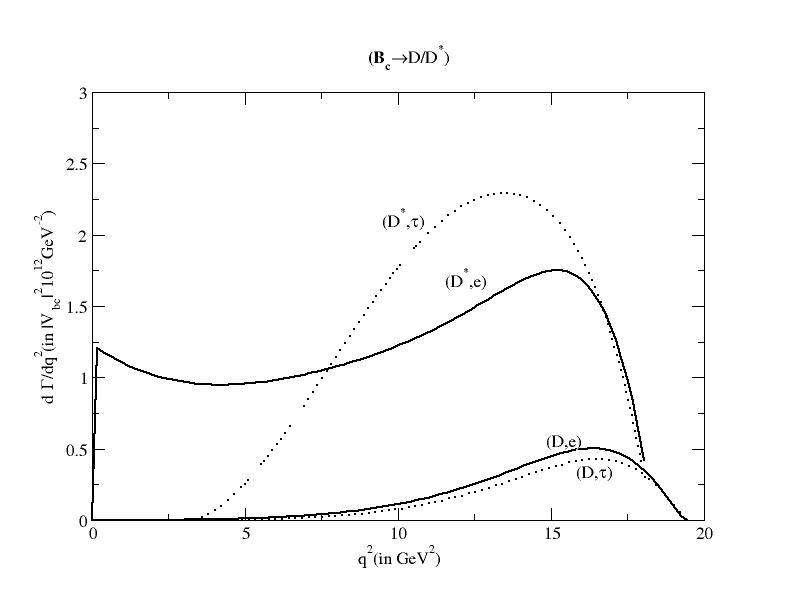}
  	\caption{$q^2-$spectrum of s.l. decay rates for $B_c\to D$ and $B_c\to D^*$ decays.}
  	\label{fig 9}
  \end{figure}

 	  In FIG.\ref{fig 8} and  \ref{fig 9}, we plot the total $q^2$-spectra: $\frac{d\Gamma}{dq^2}$ for the s.l. decays $B_c\to \eta_c(J/\psi)$ and $B_c\to D(D^*)$, respectively in their $e^-$ and $\tau^-$ modes. For $B_c\to \eta_c$ decay in $e^-$ mode, the $q^2$-spectra increases rapidly in small $q^2$-region near $q^2\to 0$ to a peak value, and then it flattens with a slow rise to another peak at $q^2\approx 6.25\ GeV^2$. Thereafter it decreases to zero at $q^2\approx 11\ GeV^2$. However, in $\tau$-mode, the spectra  originate at $q^2\approx 2\ GeV^2$ and are reduced in comparison with $e^-$ mode spectra within the range $2\le q^2\le 8\ GeV^2$ beyond which it dominates over the $e^-$ spectra contrary to the phase-space expectations. For $B_c\to J/\psi$ decay in its $e-$mode, we find a sharper rise of spectra to a peak value near $q^2\to 0$. This is followed by a further rise of spectra developing a prominent shoulder at $q^2\approx 6.25\ GeV^2$. Beyond $6.25\ GeV^2$, there occurs a sharp fall of the spectra. In contrast to $e^-$ mode, there occurs a considerable reduction of spectra in $\tau^-$ mode in the physical kinematic range of $3\le q^2\le 10\ GeV^2$. For $B_c\to D$ decay in its $e^-$ mode, $q^2$ spectra rise very slowly from $q^2\to 0$ to a peak value at $q^2\approx 16.5\ GeV^2$ and then fall to zero value at $q^2\approx 19.5\ GeV^2$. In its $\tau^-$-mode the spectra are uniformly reduced throughout. For the $B_c\to D^*$ case in $e^-$-mode there occurs a sharp rise of spectra near $q^2\to 0$ to a peak value after which it falls to a minimum value at $q^2\to 3.5\ GeV^2$. Thereafter it witnesses further rise till a second shoulder appears at $q^2\approx 16\ GeV^2$, beyond which there occurs a sharp fall of spectra in the high $q^2$ region. However, in its $\tau$ mode, the spectra are reduced in the low $q^2$ region but it overtakes the $e$ mode spectra in the high $q^2$ region beyond $7.5\ GeV^2$ contrary to the naive phase space expectations.

 	  \begin{table}[!hbt]
 	  	\renewcommand{\arraystretch}{1}
 	  	\centering
 	  	\setlength\tabcolsep{0.5pt}
 	  	\caption{Helicity rates(in $10^{-15}$ GeV) of semileptonic $B_c$-meson decays into charmonium and charm-meson state:}
 	  	\label{tab1}
 	  	\begin{tabular}{|c|c|c|c|c|c|c|c|c|c|}
 	  		\hline Decay mode &$U$ & $\tilde{U}$ & $L$ & $\tilde{L}$ &$P$&$S$ & $\tilde{S}$&$\tilde{SL}$ & $\Gamma$ \\
 	  		\hline${B^-_c}\to \eta_c e^-\nu_e$ & & & 4.844 & 4.432$10^{-7}$& & &15.397$10^{-7}$&4.712$10^{-7}$&4.844\\
 	  		
 	  		\hline$B_c\to\eta_c \tau^-\nu_\tau$ & & & 0.756 & 0.172 && &1.194 &0.253&2.122\\
 	  		
 	  		\hline${B^-_c}\to J/\psi e^-\nu_e$ &18.634&6.052$\times10^{-7}$ &16.283&27.813$\times 10^{-7}$&8.368 &1.188 &66.653$\times 10^{-7}$&22.856$\times 10^{-7}$&34.918\\
 	  		
 	  		\hline${B^-_c}\to J/\psi \tau^-\nu_\tau$ & 3.823 &0.846 &1.922& 0.437&1.704 &0.614 &0.307&0.197&7.336\\
 	  		
 	  		\hline${B^-_c}\to D e^-\nu_e$ &  &  &0.047&4.611$\times 10^{-10}$& & &1.072$\times 10^{-9}$&4.038$\times 10^{-10}$&0.047\\
 	  		
 	  		\hline${B^-_c}\to D\tau^-\nu_\tau$ &  &&0.028&0.003& &&0.007&0.0027&0.038\\
 	  		
 	  		\hline${B^-_c}\to D^* e^-\nu_e$ &0.2439&4$\times 10^{-9}$&0.078&7.760$\times 10^{-9}$&0.169&0.081&4.092$\times 10^{-8}$&3.648$\times 10^{-9}$&0.322\\
 	  		
 	  		\hline${B^-_c}\to D^*\tau^-\nu_\tau$ &0.113&0.015&0.0156&0.0021&0.092&0.046&0.151&0.0094&0.297\\
 	  		\hline
 	  	\end{tabular}
 	  \end{table}

 	  In TABLE \ref{tab1} we list our results for the integrated partial helicity rates: $\Gamma_i(i=U,L,P)$ and $\tilde{\Gamma}_i(i=U,L,S,SL)$ as well as the total decay rates. The partial tilde rates for each decay process in the $e^-$-mode are found tiny as expected  from Eq.(23) and can therefore be neglected. But corresponding rates in $\tau^-$-mode obtained comparable to the tilde rates, can hardly be neglected. Considering contribution from individual helicity rates we predict the decay rates for each process in its $e^-$ as well as $\tau^-$ mode. Like all other model predictions, our predicted decay rates in $\tau^-$ modes are, in general, found smaller than that in $e^-$ modes. For $B_c\to J/\psi$ transition, our predicted decay rate in $\tau^-$-mode is suppressed compared to its corresponding value in $e^-$ mode by a factor of $\sim$5; whereas for $B_c\to \eta_c$ transition, the suppression is by a factor of $\sim$2. For the CKM suppressed $B_c\to D(D^*)$ transitions, the $\tau^-$-modes are down only marginally over corresponding in $e^-$-mode.\\
 	  
 	  With the central value of $B_c$-meson lifetime: $\tau_{B_c}=0.507\ ps$ and our predicted decay rates(TABLE \ref{tab1}), we predict the branching fractions(B.F) for s.l. $B_c$-meson decays into charmonium and charm meson states and our results are listed in TABLE \ref{tab2} in comparison with other model predictions. The predictions, in this sector obtained from different theoretical approaches, are in the same order of magnitude. Like all other model predictions, we find that our predicted B.F for s.l. $B_c\to J/\psi$ transition is the largest of all and comparable to the model predictions \cite{A10,A23,A46}. For $B_c\to \eta_c$ transition, our predicted branching fraction although is down by a factor of $\sim$2 compared to that of \cite{A10,A23,A46}, is comparable to that of \cite{A44,A49}. 
 	  \begin{table}
 	  	\centering
 	  	\caption{Branching ratios(in\%) of semileptonic $B_c$ decays into ground state charmonium and charm meson state:}
 	  	\label{tab2}
 	  	\begin{tabular}{|c|c|c|c|c|c|c|c|c|c|c|c|c|}
 	  		\hline Decay mode&This work&\cite{A24}&\cite{A46}&\cite{A23}&\cite{A51,A52}&\cite{A10}&\cite{A44}&\cite{A49}&\cite{A24}&\cite{A11,A12}&\cite{A57}&\cite{A58}\\
 	  		\hline$B_c\to \eta_ce\nu$&0.37&0.83&0.81&0.98&0.75&0.97&0.59&0.44&0.95&0.86&0.162&0.45\\
 	  		$B_c\to \eta_c\tau \nu$&0.16&0.27&0.22&0.27&0.23&-&0.20&0.14&0.24&-&-&-\\
 	  		\hline $B_c\to J/\psi e\nu$&2.68&2.19&2.07&2.30&1.9&2.35&1.20&1.01&1.67&2.33&1.67&1.37\\
 	  		$B_c\to J/\psi \tau\nu$&0.56&0.61&0.49&0.59&0.48&-&0.34&0.29&0.40&-&-&-\\
 	  		\hline$B_c\to De\nu$&0.0037&-&0.0035&0.018&-&0.004&0.004&0.0032&0.0033&-&-&-\\
 	  		$B_c\to D\tau \nu$&0.0029&-&0.0021&0.0094&0.002&-&-&0.0022&0.0021&-&-&-\\
 	  		\hline $B_c\to D^* e\nu$&0.0251&-&0.0038&0.034&-&0.018&0.018&0.011&0.006&-&-&-\\
 	  		$B_c\to D^*\tau\nu$&0.0230&-&0.0022&0.019&0.008&-&-&0.006&0.0034&-&-&-\\
 	  		\hline
 	  	\end{tabular}
 	  \end{table}
 	  
 	    As expected the B.Fs for $B_c\to D(D^*)$ transitions induced by $b\to u$ transition at the quark level, are down in comparison with that of $B_c\to \eta_c(J/\psi)$. Our result for $B_c\to D$ transition agrees with those of \cite{A25,A46,A49}. For $B_c\to D^*$transition our result is in reasonable agreement with those of \cite{A10,A44} though it is larger in comparison with most other model predictions by a factor of $\sim~6$ and $\sim ~10$ in their $e^-$- and $\tau^-$-modes, respectively.\\
 	    
 	 Considering contribution from different partial helicity rates, we evaluate the quantities of interest: the forward-backward asymmetry $'A_{FB}'$ and  asymmetry parameter $\alpha^*$. The results are shown in TABLE.\ref{tab3}. For the decay into spin 0 state, $A_{FB}$ is proportional to the helicity amplitude $\tilde{SL}$, which is also tiny in $e-$mode but non-negligible for transitions in the $\tau-$mode.    
 	   
 	   \begin{table}[hbt!]
 	  	\centering
 	  	\caption{Forward-backward asymmetry $A_{FB}$ and the  asymmetry parameter $\alpha^*$ }
 	  	\label{tab3}
 	  	\begin{tabular}{|c|c|c|c|}
 	  		\hline Decay process &$A_{FB}(l^-)$&$A_{FB}(l^+)$&$\alpha^*$\\
 	  		\hline$B_c\to \eta_ce\nu$&2.049$\times10^{-7}$&2.049$\times10^{-7}$&\\
 	  		$B_c\to \eta_c\tau \nu$&0.357&0.357&\\
 	  		\hline $B_c\to J/\psi e\nu$&0.180&0.180&-0.27\\
 	  		$B_c\to J/\psi \tau\nu$&0.093&-0.255&-0.066\\
 	  		\hline$B_c\to De\nu$&2.55$\times10^{-8}$&2.55$\times10^{-8}$&\\
 	  		$B_c\to D\tau \nu$&0.210&0.210&\\
 	  		\hline $B_c\to D^* e\nu$&0.394&-0.394&0.22\\
 	  		$B_c\to D^*\tau\nu$&0.137&-0.328&-0.45\\
 	  		\hline
 	  		
 	  	\end{tabular}
 	  \end{table}

 For decays into spin 1 states, we obtained $A_{FB}(e^-)=-A_{FB}(e^+)$ and $A_{FB}(\tau^-)\ne -A_{FB}(\tau^+)$ as shown in TABLE \ref{tab3}. This is comprehensible by looking at the expression of $A_{FB}$ in Eq(25). The transverse and longitudinal pieces of $J/\psi $ in $B_c\to J/\psi l\nu$ are found almost equal in $e^-$ mode, but the transverse component dominates over the longitudinal part for this transition  in its $\tau-$mode by a factor of $\sim$2. However, in the $B_c\to D^*$ transition, the transverse component of the partial helicity rates also dominates over the longitudinal part by a factor of $\sim$3 and $\sim$7, respectively in their $e^-$ and $\tau^-$ modes. To determine the transverse and longitudinal component of the final state vector mesons in $B_c\to J/\psi$ and $B_c\to D^*$ transitions, we calculate the asymmetry parameter $\alpha^*$.  For $B_c\to J/\psi$ transition, the asymmetry parameter is found close to $-27\%$ in the $e^-$ mode and $-7\%$ in its $\tau$ mode. Our predicted $\alpha^*$ for $B_c\to D^*$ transitions is found close to $+22\%$ in the $e^-$ mode while it is found as low as $-45\%$ in its $\tau^-$ mode. This is because for the later case in the $e^-$mode the transverse component of the helicity amplitudes provides a significant contribution compared to a tiny contribution coming from both the longitudinal and scalar parts. On the other hand, in the $\tau^-$-mode, the scalar flip part $\tilde{S}$ of the helicity amplitude dominates over the rest part and contributes destructively resulting in a highly suppressed asymmetry parameter as low as $-45\%$.
 	
 	    \begin{table}[hbt!]
 	  	\centering
 	  	\caption{Ratios of Branching Fractions for Semileptonic $B_c-$decays}
 	  	\label{tab4}
 	  	\begin{tabular}{|c|c|c|c|c|}
 	  		\hline Ratio of Branching Fractions($R$)&This work&\cite{A25}&\cite{A46}&\cite{A49}\\
 	  		\hline $R_{\eta_c}=\frac{{\cal B}(B_c\to \eta_c l\nu)}{{\cal B}(B_c\to \eta_c \tau \nu)}$&2.312 &3.96&3.68&3.2\\
 	  		\hline $R_J/\psi=\frac{{\cal B}(B_c\to J/\psi l\nu)}{{\cal B}(B_c\to J\psi \tau \nu)}$&4.785&4.18&4.22&3.4\\
 	  		\hline$R_D=\frac{{\cal B}(B_c\to D l\nu)}{{\cal B}(B_c\to D\tau \nu)}$&1.275&1.57&1.67&1.42\\
 	  		$R_{D^*}=\frac{{\cal B}(B_c\to D^* l\nu)}{{\cal B}(B_c\to D^* \tau \nu)}$&1.091&1.76&1.72&1.66\\
 	  		\hline
 	  	\end{tabular}
 	  \end{table}
 	  
 	  Finally, we evaluate the observable $'R'$ and our results as listed in TABLE.\ref{tab4}, are comparable to other SM predictions which highlight the puzzle in flavor physics and failure of lepton flavor universality. This hints at new physics beyond SM for the explanation of the observed deviation of observable $'R'$ from the SM predictions.\\
 	  
 	  \section{Summary and Conclusion:}
 	  In this paper, we analyze exclusive $B_c$-meson decays into charmonium and charm meson ground states in the framework of the relativistic independent quark(RIQ) model based on an average flavor-independent confining potential in an equally mixed scalar-vector harmonic form. The invariant form factors representing decay amplitudes are extracted from the overlapping integral of meson wave functions derivable from the model dynamics. Since our main objective here is to evaluate the lepton mass effects on the semileptonic $B_c$-meson decays, we analyze $B_c\to \eta_c(J/\psi)l\nu $ and $B_c\to D(D^*)l\nu$ in their $e^-$ and $\tau^-$ modes separately. For this, we study the $q^2$-dependence of relevant physical quantities such as the invariant form factors, helicity form factors, partial helicity rates, and the total $q^2$-spectra for all decay processes analyzed in the present work.\\
 	  
 	  Considering contribution from different partial helicity rates: $\frac{d\Gamma_i}{dq^2}\ (i=U,L,P)$ and $\frac{d\tilde{\Gamma}_i}{dq^2}\ (i=U,L,S,SL)$, the total $q^2$-spectrum $\frac{d\Gamma}{dq^2}$ is obtained for each  decay process, from which we predict the decay rates/branching fractions of the semileptonic $B_c$-meson decays into their $e^-$ as well as $\tau^-$-modes. Our prediction on the decay rates/ branching fractions for decay processes is found in overall agreement with other SM predictions. We find that the decay rate for $B_c\to J/\psi $ transition is the largest of all. As expected, the decay rates for $B_c\to \eta_c(J/\psi)$ transitions induced by the quark level $b\to c$ transition dominate over those for $B_c\to D(D^*)$ transitions induced by the quark level $b\to u$ transition in their respective $e^-$ as well as $\tau^-$ modes. We also find that the $\tau^-$ modes for all decays are, in general, suppressed in comparison with their corresponding $e^-$ modes.\\
 	  
 	  Using appropriate helicity rates we evaluate two quantities of interest:(1)Forward-backward asymmetry ‘$A_{FB}$’ and the asymmetry parameter ‘$\alpha^*$’. $A_{FB}$ in the present analysis is found negligible for transitions into spin 0 state in their $e^-$-mode but non-negligible in $\tau^-$-mode, as expected. For transition into spin 1 state, we also find $A_{FB}(e^-)=-A_{FB}(e^+)$ and $A_{FB}(\tau^-)\ne -A_{FB}(\tau ^+)$. The asymmetry parameter $\alpha^*$, which determines the transverse and longitudinal components of the final vector meson state for $B_c\to J/\psi$ and $B_c\to D^*$ transitions, have been evaluated. Our predicted $\alpha^*$ for $B_c\to J/\psi$ transition is close to $-27\%$ in the $e^-$ mode and $-7\%$ in its $\tau^-$-mode whereas for $B_c\to D^*$ transition it is found close to $+22\%$ in $e^-$-mode and as low as $-45\%$ in $\tau^-$-mode.\\
 	  
 	  Finally, we evaluate the observable $'R'$, which corresponds to the ratios of branching fractions for transitions in $e^-$ modes concerning their corresponding values in $\tau^-$ modes. Our predicted observable $'R'$ is found comparable to other SM predictions, which highlights the observed deviation of observable $'R'$ from corresponding SM predictions. This is indicative of the failure of lepton flavor universality that hints at new physics beyond SM to explain the anomaly between the observed data and SM predictions. Our predictions in this sector can be useful to identify the $B_c$-channels characterized by clear experimental signature. Considering the possibility of high statistics $B_c$-events expected to yield up to $10^{10}$ events per year at the Tevatron and LHC, semileptonic $B_c$-meson decays into charmonium and charm mesons in their ground as well as excited states offer a fascinating area for future research. 
 	  \begin{acknowledgements} 
 	  	The library and computational facilities provided by the authorities of Siksha 'O' Anusandhan Deemed to be University, Bhubaneswar, 751030, India are duly acknowledged.
 	  \end{acknowledgements}

 \appendix
 \section{CONSTITUENT QUARK ORBITALS AND MOMENTUM PROBABILITY AMPLITUDES}\label{app}
 
 In the RIQ model, a meson is picturized as a color-singlet assembly of a quark and an antiquark independently confined by an effective and average flavor independent potential in the form:
 $U(r)=\frac{1}{2}(1+\gamma^0)(ar^2+V_0)$ where ($a$, $V_0$) are the potential parameters. It is believed that the zeroth-order quark dynamics  generated by the phenomenological confining potential $U(r)$ taken in equally mixed scalar-vector harmonic form can provide an adequate tree-level description of the decay process being analyzed in this work. With the interaction potential $U(r)$ put into the zeroth-order quark lagrangian density, the ensuing Dirac equation admits a static solution of positive and negative energy as: 
 \begin{eqnarray}
 	\psi^{(+)}_{\xi}(\vec r)\;&=&\;\left(
 	\begin{array}{c}
 		\frac{ig_{\xi}(r)}{r} \\
 		\frac{{\vec \sigma}.{\hat r}f_{\xi}(r)}{r}
 	\end{array}\;\right)U_{\xi}(\hat r)
 	\nonumber\\
 	\psi^{(-)}_{\xi}(\vec r)\;&=&\;\left(
 	\begin{array}{c}
 		\frac{i({\vec \sigma}.{\hat r})f_{\xi}(r)}{r}\\
 		\frac{g_{\xi}(r)}{r}
 	\end{array}\;\right){\tilde U}_{\xi}(\hat r)
 \end{eqnarray}
 where, $\xi=(nlj)$ represents a set of Dirac quantum numbers specifying 
 the eigen-modes;
 $U_{\xi}(\hat r)$ and ${\tilde U}_{\xi}(\hat r)$
 are the spin angular parts given by,
 \begin{eqnarray}
 	U_{ljm}(\hat r) &=&\sum_{m_l,m_s}<lm_l\;{1\over{2}}m_s|
 	jm>Y_l^{m_l}(\hat r)\chi^{m_s}_{\frac{1}{2}}\nonumber\\
 	{\tilde U}_{ljm}(\hat r)&=&(-1)^{j+m-l}U_{lj-m}(\hat r)
 \end{eqnarray}
 With the quark binding energy $E_q$ and quark mass $m_q$
 written in the form $E_q^{\prime}=(E_q-V_0/2)$,
 $m_q^{\prime}=(m_q+V_0/2)$ and $\omega_q=E_q^{\prime}+m_q^{\prime}$, one 
 can obtain solutions to the resulting radial equation for 
 $g_{\xi}(r)$ and $f_{\xi}(r)$in the form:
 \begin{eqnarray}
 	g_{nl}&=& N_{nl} (\frac{r}{r_{nl}})^{l+l}\exp (-r^2/2r^2_{nl})
 	L_{n-1}^{l+1/2}(r^2/r^2_{nl})\nonumber\\
 	f_{nl}&=& N_{nl} (\frac{r}{r_{nl}})^{l}\exp (-r^2/2r^2_{nl})\nonumber\\
 	&\times &\left[(n+l-\frac{1}{2})L_{n-1}^{l-1/2}(r^2/r^2_{nl})
 	+nL_n^{l-1/2}(r^2/r^2_{nl})\right ]
 \end{eqnarray}
 where, $r_{nl}= a\omega_{q}^{-1/4}$ is a state independent length parameter, $N_{nl}$
 is an overall normalization constant given by
 \begin{equation}
 	N^2_{nl}=\frac{4\Gamma(n)}{\Gamma(n+l+1/2)}\frac{(\omega_{nl}/r_{nl})}
 	{(3E_q^{\prime}+m_q^{\prime})}
 \end{equation}
 and
 $L_{n-1}^{l+1/2}(r^2/r_{nl}^2)$ etc. are associated Laguerre polynomials. The radial solutions yields an independent quark bound-state condition in the form of a cubic equation:
 \begin{equation}
 	\sqrt{(\omega_q/a)} (E_q^{\prime}-m_q^{\prime})=(4n+2l-1)
 \end{equation}
 The solution of the cubic equation provides the zeroth-order binding energies of 
 the confined quark and antiquark for all possible eigenmodes.
 
 In the relativistic independent particle picture of this model, the constituent quark 
 and antiquark are thought to move independently inside the $B_c$-meson bound state 
 with momentum $\vec p_b$ and $\vec p_c$, respectively. Their momentum probability 
 amplitudes are obtained in this model via momentum projection of respective quark orbitals (A1) in the following forms:
 
 For ground state mesons:($n=1$,$l=0$)
 \begin{eqnarray}
 	G_b(\vec p_b)&&={{i\pi {\cal N}_b}\over {2\alpha _b\omega _b}}
 	\sqrt {{(E_{p_b}+m_b)}\over {E_{p_b}}}(E_{p_b}+E_b)\nonumber\\
 	&&\times\exp {(-{
 			{\vec {p_b}}^2\over {4\alpha_b}})}\nonumber\\
 	{\tilde G}_c(\vec p_c)&&=-{{i\pi {\cal N}_c}\over {2\alpha _c\omega _c}}
 	\sqrt {{(E_{p_c}+m_c)}\over {E_{p_c}}}(E_{p_c}+E_c)\nonumber\\
 	&&\times\exp {(-{
 			{\vec {p_c}}^2\over {4\alpha_c}})}
 \end{eqnarray}

 The binding energy of constituent quark and antiquark in the parent and daughter meson in their ground-states are obtained by solving respective cubic equations representing appropriate bound state conditions(A5).


\newpage 
 
 
 	  
 	  
 	  
	 
	  
  
	\begin{thebibliography}{90}
		\bibitem{A1}
	 F.Abe {\it et al.}, CDF Collaboration, {\it Phys.Rev.}  D {\bf 58}, 112004 (1998);	 F.Abe {\it et al.}, CDF Collaboration, {\it Phys. Rev. Lett.}{\bf 81}, 2432 (1998).
	 \bibitem{A2}
	 A. Abulencia {\it et.al., Phys.Rev.Lett.} {\bf 97},012002(2006).
	 \bibitem{A3}
	 V. Abazov {\it et.al.},{\it Phys.Rev.Lett.} {\bf 102}, 092001(2009).
		\bibitem{A4}
		T. A. Altoner {\it et.al.}, {\it Phys.Rev.Lett.} {\bf 100}, 182008 (2008).
		\bibitem{A5}
		V.M. Abazov {\it et. al.}, {\it Phys.Rev.Lett.}, {\bf 101}, 012001 (2008).
		\bibitem{A6} 
			R. Aajj {\it et al.}(LHCb Collaboration){\it Eur. Phys. J. C.} {\bf 74}, 2839 (2014).
		\bibitem{A7}
		G. Aad {\it et. al.}(ATLAS Collaboration), {\it Phys. Rev. Lett.},{\bf 113}, 212004 (2014).
		\bibitem{A8}
		D. Lucchesi, in Proceedings of the ICHEP04, BEIJING, 2004(to be published). See also S.Towers. Do Note No. 4539-CONF.2004
		\bibitem{A9}
		S. D'Auria, in Proceedings of the "Fermilab Joint Experiment Theoretical Seminar",2004.
	\bibitem{A10}
	C. H. Chang and Y. Q. Chen, {\it Phys. Rev.} D {\bf 49}, 3399 (1994).
\bibitem{A11}
A. Abd El-Hady, J. H. Munoz, and J. P. Vary, {\it Phys. Rev.} D {\bf 62},014019 (2000).
\bibitem{A12}
J. F. Liu and K. T. Chao, {\it Phys.Rev.}D {\bf 56},4133 (1997).
\bibitem{A13}
A. Y. Anisimov, P. Y. Kulikov, I. M. Narodetsky, and K. A. Ter-Martirosian, Yad. Fiz.{\bf 62}, 1868(1999);[{\it Phys. At. Nucl.} {\bf 62}, 1739 (1999)].
\bibitem{A14}
W. Wang, Y. L. Shen, and C. D. Lu, {\it Phys. Rev.} D {\bf 79}, 054012 (2009).
\bibitem{A15}
V. V. Kiselev, A.K. Likhoded, and A. I. Onishchenko, {\it Nucl. Phys.} {\bf B569},473(2000).
\bibitem{A16}
V. V. Kiselev, A. E. Kovalsky, and A. K. Likhoded, {\it Nucl. Phys.} {\bf B585}, 353 (2000).
  \bibitem{A17}
  V. V. Kiselev, arXiv:hep-ph/0211021.
  \bibitem{A18}
  D. Ebert, R. N. Faustov, and V. O. Galkin, {\it Phys. Rev.} D {\bf 68}, 094020 (2003).
  \bibitem{A19}
  E. Hernandez, J. Nieves, and J. M. Verde-Velasco, {\it Phys. Rev.} D {\bf 74}, 074008 (2006).
  \bibitem{A20}
  R. Dhir and R. C. Verma, {\it Phys. Rev.} D {\bf 79},034004 (2009).
  \bibitem{A21}
  W. F. Wang, Y. Y. Fan and Z. J. Xiao, {\it Chin. Phys.} C {\bf 37}
  093102 (2013).
  \bibitem{A22}
  Z. Rui, H. Li, G. X. Wang and Y. Xiao,{\it Eur. Phys. J.} C {\bf 76} 564 (2016).
  \bibitem{A23}
  M. A. Ivanov, J. G. K{$\ddot{o}$}rner, and P. Santorelli, {\it Phys. Rev.} D {\bf 63}, 074010 (2001).
  \bibitem{A24}
  M. A. Ivanov,J. G. K$\ddot{o}$rner, and P. Santorelli, {\it Phys. Rev.} D {\bf 71}, 094006(2005); {\bf 75}, 019901(E)(2007).
   \bibitem{A25}
   A. Issadykov, M.A. Ivanov, and G. Nurbakova, EPJ Web Conf. {\bf 158}, 03002 (2017).
  \bibitem{A26}
  C. T. Tran, M. A. Ivanov, J. G. Korner and P. Santorelli, {\it Phys. Rev.}D {\bf 97}, 054014 (2018).
  \bibitem{A27}
 B. Colquhoun, C. Davis, J. Kopnen, A. Lytle and C. McNeil (HPQCD Collaboration), {\it Proc. Sci} LATTICE2016 (2016) 281.
  \bibitem{A28}
  N. Barik, B. K. Dash, {\it Phys. Rev.} D {\bf 33}, 1925 (1986); N. Barik, B. K. Dash, P. C. Dash{\bf 29} 543 (1987); N. Barik, P. C. Dash, {\it Phys. Rev.} D {\bf 47}, 2788 (1993).
\bibitem{A29}
N. Barik, P. C. Dash, A. R. Panda, {\it Phys. Rev.} D {\bf 46}, 3856 (1992); N. Barik, P. C. Dash, {\it Phys. Rev.} D {\bf 49}, 299 (1994); M. Priyadarsini, P. C. Dash, S. Kar, S. P. Patra, N. Barik, {\it Phys. Rev.} D {\bf 94}, 113011 (2016); N. Barik, P. C. Dash, {\it Mod. Phys. Lett.} A {\bf 10}, 103 (1995); N. Barik, S. Kar, P. C. Dash, {\it Phys.Rev.} D {\bf 57}, 405 (1998); N. Barik, Sk. Naimuddin, S. Kar, P. C. Dash, {\it Phys. Rev.} D {\bf 63}, 014024 (2000).
\bibitem{A30}
N. Barik, P. C. Dash, A. R. Panda, {\it Phys. Rev.} D {\bf 47}, 1001 (1993); N. Barik, P. C. Dash, {\it Phys. Rev.} D {\bf 47}, 2788 (1993); N. Barik, Sk. Naimuddin, P. C. Dash, S. Kar, {\it Phys. Rev.} D {\bf 77}, 014038 (2008); N. Barik, Sk. Naimuddin, P. C. Dash, S. Kar, {\it Phys. Rev.} D {\bf 77}, 014038 (2008); N. Barik, Sk. Naimuddin, P. C. Dash, S. Kar, {\it Phys. Rev.} D {\bf 78}, 114030 (2008); N. Barik, Sk. Naimuddin, P. C. Dash, {\it Mod. Phys.} A{\bf 24},2335 (2009).
\bibitem{A31}
N, Barik, P. C. Dash, {\it Phys. Rev.} D {\bf 63}, 114002 (2001); S. Kar, P. C. Dash, M. Priyadarsini, Sk Naimuddin, N. Barik, {\it Phys. Rev.} D {\bf 88}, 094014 (2013); N.Barik, Sk. Naimuddin, P. C. Dash, S. Kar, {\it Phys. Rev.} D {\bf 80}, 014004 (2009); Sk. Naimuddin, S. Kar, M. Priyadarshini, N. Barik, P. C. Dash, {\it Phys. Rev.} D {\bf 86}, 094028 (2012).
\bibitem{A32}
N. Barik, P.C. Dash, {\it Phys. Rev.} D {\bf 53},1366 (1996).
\bibitem{A33}
N. Barik, S. K. Tripathy, S. Kar, P. C. Dash, {\it Phys. Rev.} D {\bf 56}, 4238 (1997).
\bibitem{A34}
N. Barik, Sk. Naimuddin, P. C. Dash, S. Kar, {\it Phys. Rev. } D {\bf 80}, 074005 (2009).
\bibitem{A35}
Sonali Patnaik, Lopamudra Nayak, P. C. Dash, S. Kar, N. Barik, {\it Eur. Phys. J. Plus} (2000) 135:936.
\bibitem{A36}
 J.P.Lees, {\it et al.}, BaBar Collaboration, {\it Phys. Rev. Lett.} 109(2012) 101802, https://doi.org/10.1103/PhysRevLett.109.101802, arXiv:1205.5442 [hep-ex].
 \bibitem{A37}
 J.P. Lees, {\it et al.}, BaBar Collaboration,{\it Phys. Rev.} D {\bf 88} (7) (2013) 072012, https://
 doi.org/10.1103/PhysRevD.88.072012, arXiv:1303.0571 [hep-ex].
 \bibitem{A38}
 M. Huschle, {\it et al.}, Belle Collaboration, {\it Phys. Rev.} D {\bf 92} (7) (2015) 072014,
 https://doi.org/10.1103/PhysRevD.92.072014, arXiv:1507.03233 [hep-ex].
\bibitem{A39}
Y. Sato, {\it et al.}, Belle Collaboration, {\it Phys. Rev.} D {\bf 94} (7) (2016) 072007, https://
doi.org/10.1103/PhysRevD.94.072007, arXiv:1607.07923 [hep-ex].
\bibitem{A40}
S. Hirose, {\it et al.}, Belle Collaboration, {\it Phys. Rev. Lett.} 118 (21) (2017) 211801,
https://doi.org/10.1103/PhysRevLett.118.211801, arXiv:1612.00529 [hep-ex]
\bibitem{A41}
R. Aaij, {\it et al.}, LHCb Collaboration, {\it Phys. Rev. Lett.} 115 (11) (2015) 111803,
https://doi.org/10.1103/PhysRevLett.115.111803, arXiv:1506.08614 [hep-ex];{\it Phys. Rev. Lett.} 115 (15) (2015) 159901, https://doi.org/10.1103/
PhysRevLett.115.159901 (Erratum).
\bibitem{A42}
LHCb Collaboration at FPCP2017.
\bibitem{A43}
R. Aaij, {\it et al.}, LHCb Collaboration, arXiv:1711.05623 [hep-ex].
\bibitem{A44}
 A. Y. Anisimov, I. M. Narodetsky, C. Semay and B. Silvestre-Brac, {\it Phys. Lett.} B {\bf 452}, 129
(1999) [arXiv:hep-ph/9812514]; A. Y. Anisimov, P. Y. Kulikov, I. M. Narodetsky and K. A. Ter-Martirosian, {\it Phys. Atom. Nucl.} {\bf 62}, 1739 (1999) [Yad. Fiz. 62, 1868 (1999)] [arXiv:hep-ph/9809249].
\bibitem{A45}
V.V. Kiselev, arXiv:hep-ph/0211021.
\bibitem{A46}
M.A. Ivanov, J.G. Korner, P. Santorelli, {\it Phys. Rev.} D {\bf 73} (2006) 054024, https://
doi.org/10.1103/PhysRevD.73.054024, arXiv:hep-ph/0602050.
\bibitem{A47}
E. Hernandez, J. Nieves, J.M. Verde-Velasco, Phys. Rev. D 74 (2006) 074008,
https://doi.org/10.1103/PhysRevD.74.074008, arXiv:hep-ph/0607150.
\bibitem{A48}
 R. Watanabe, arXiv:1709.08644 [hep-ph].

 \bibitem{A49}
 W. F. Wang, Y. Y. Fan and Z. J. Xiao, {\it Chin. Phys.} C {\bf 37} (2013) 093102; doi10.1088/1674-1137/37/9/093102[arXiv:1212.5903[hep-ph]]
 \bibitem{A50}
P.A. Zyla et al. (Particle Data Group), Prog. Theor. Exp. Phys. 2020, 083C01 (2020).
 \bibitem{A51}
J. G. K$\ddot{o}$rner and G. A. Schuler, {\it Z. Phys.} C {\bf 46}, 93 (1990).
\bibitem{A52}
J. G. K$\ddot{o}$rner, J. H. K$\ddot{u}$hn, and H. Schneider, {\it Phys. Lett.} {\bf 120B}, 444 (1983).
\bibitem{A53}
J. G. K$\ddot{o}$rner, J. H. K$\ddot{u}$hn, M. Krammer, and H. Schneider,{\it Nucl. Phys.} {\bf B229}, 115 (1983).
\bibitem{A54}
B. Margolis and R. R. Mendel, {\it Phys. Rev.} D {\bf 28}, 468(1983).

 \bibitem{A55}
 V. V. Kiselev, A. E. Kovalsky, and A. K. Likhoded, {\it Nucl. Phys.} {\bf B } 585, 353 (2000) [arXiv:hep-ph/0002127]; arXiv:hep-ph/0006104.

 \bibitem{A56}
 V. V. Kiselev, arXiv:hep-ph/0211021.

 \bibitem{A57}
 P. Colangelo and F. De Fazio, {\it Phys. Rev.} {\bf D} 61, 034012
 (2000).
 \bibitem{A58}
  D. Ebert, R. N. Faustov, and V. O. Galkin, {\it Phys. Rev.} {\bf D} 68,
 094020 (2003).

	\end{thebibliography}
\end{document}